\documentclass[a4paper,fleqn,usenatbib]{mnras}

\usepackage{newtxtext,newtxmath}

\usepackage[T1]{fontenc}
\usepackage{ae,aecompl}

\usepackage[switch]{lineno}

\usepackage{graphicx}	
\usepackage{amsmath}	
\usepackage{amssymb}	
\usepackage{etoolbox}
\makeatletter
\patchcmd\@combinedblfloats{\box\@outputbox}{\unvbox\@outputbox}{}{%
   \errmessage{\noexpand\@combinedblfloats could not be patched}%
}%
 \makeatother

\newcommand{\ltsima}{$\; \buildrel < \over \sim \;$}
\newcommand{\simlt}{\lower.5ex\hbox{\ltsima}}
\newcommand{\gtsima}{$\; \buildrel > \over \sim \;$}
\newcommand{\simgt}{\lower.5ex\hbox{\gtsima}}

\newcommand{\cgs}{${\rm erg~cm}^{-2}~{\rm s}^{-1}$} 
 
\newcommand{\lum}{\rm erg~s$^{-1}$}

\newcommand{\pn}{\par\noindent}

\def\lesssim{\mathrel{\hbox{\rlap{\hbox{\lower4pt\hbox{$\sim$}}}\hbox{$<$}}}}
\def\gtrsim{\mathrel{\hbox{\rlap{\hbox{\lower4pt\hbox{$\sim$}}}\hbox{$>$}}}}

\def\arcsec{\hbox{$^{\prime\prime}$}}
\def\micron{\hbox{$\mu$m}}

\def\aox{$\alpha_{\rm ox}$}

\def\ab1450{$AB_{1450(1+z)}$}

\def\xray{\hbox{X-ray}}

\def\cii{\hbox{C\ {\sc ii}}}
\def\oiii{\hbox{[O\ {\sc iii}}]}

\def\cii{\hbox{C\ {\sc ii}}}
\def\civ{\hbox{C\ {\sc iv}}}

\def\nh{$N_{\rm H}$}
\def\09104{IRAS~09104$+$4109}
\def\I09104{I09104}

\def\msun{M$_{\odot}$}
\def\edd_ratio{$\log\ L_{\rm bol}/L_{\rm Edd}$}

\def\l58{{$(\lambda L_{\lambda})_{\mbox{{\rm \scriptsize 5.8\micron}}}$}}
\def\lmir2{{$(\lambda L_{\lambda})_{\mbox{{\rm \scriptsize 12.3\micron}}}$}}
\def\s1{{S$_{\mbox{{\rm \scriptsize 3.6\micron}}}$}}
\def\irac2{{S$_{\mbox{{\rm \scriptsize 4.5\micron}}}$}}
\def\f3{{S$_{\mbox{{\rm \scriptsize 5.8\micron}}}$}}
\def\mic8{{S$_{\mbox{{\rm \scriptsize 8\micron}}}$}}
\def\f24{{F$_{\mbox{{\rm \scriptsize 24\micron}}}$}}

\def\chandra{{\it Chandra\/}}

\def\heao1{{\it HEAO-1\/}}

\def\alma{{\it ALMA\/}}

\def\xmm{{XMM-{\it Newton\/}}}

\def\swift{{\it Swift\/}}

\title[X-ray emission from high-redshift dual quasars]{Probing black hole accretion in quasar pairs at high redshift}

\author[C. Vignali et al.]{
C. Vignali,$^{1,2}$\thanks{E-mail: cristian.vignali@unibo.it}
E. Piconcelli,$^{3}$
M. Perna,$^{4}$
J. Hennawi,$^{5,6}$
R. Gilli,$^{2}$
A. Comastri,$^{2}$
\newauthor
G. Zamorani,$^{2}$
M. Dotti,$^{7}$
and S. Mathur$^{8,9}$
\\
$^{1}$ Dipartimento di Fisica e Astronomia, Alma Mater Studiorum, Universit\`a degli Studi di Bologna, Via Gobetti 93/2, 40129 Bologna, Italy\\
$^{2}$ INAF -- Osservatorio di Astrofisica e Scienza dello Spazio di Bologna, Via Gobetti 93/3, 40129 Bologna, Italy\\
$^{3}$ INAF -- Osservatorio Astronomico di Roma, Via Frascati 33, 00040 Monteporzio 
Catone, Roma, Italy\\
$^{4}$ INAF -- Osservatorio Astrofisico di Arcetri, Largo Enrico Fermi 5, 50125 Florence, Italy\\
$^{5}$ Department of Physics, University of California, Santa Barbara, CA 93106, USA\\
$^{6}$ Max-Planck-Institut f{\"u}r Astronomie, K{\"o}nigstuhl 17, D-69117 Heidelberg, Germany\\
$^{7}$ Dipartimento di Fisica ``G. Occhialini", Universit\`a degli Studi di Milano-Bicocca, Piazza della Scienza 3, I-20126 Milano, Italy\\
$^{8}$ Department of Astronomy, The Ohio State University, 4055 McPherson Laboratory, 140
West 18th Avenue, Columbus, OH 43210-1173, USA\\
$^{9}$ Center for Cosmology and Astro-Particle Physics (CCAPP), 191 West Woodruff Avenue, Columbus, OH 43210, USA}

\date{Accepted 2018 March 12. Received 2018 March 12; in original form 2017 August 04}

\pubyear{2017}

\begin{document}


\label{firstpage}
\pagerange{\pageref{firstpage}--\pageref{lastpage}}
\maketitle

\begin{abstract}
Models and observations suggest that luminous quasar activity is triggered by 
mergers, so it should preferentially occur in the most massive 
primordial dark matter haloes, where the frequency of mergers is expected to 
be the highest. Since the importance of galaxy mergers increases with 
redshift, we identify the high-redshift Universe as the ideal laboratory for 
studying dual AGN. 
Here we present the X-ray properties of two systems of dual quasars at 
z=3.0--3.3 selected from the SDSS DR6 at separations of 6--8 arcsec (43--65~kpc) 
and observed by \chandra\ for $\approx65$~ks each. Both members of each pair 
are detected with good photon statistics to allow us to constrain the column 
density, spectral slope and intrinsic X-ray luminosity. 
%
We also include a recently discovered dual quasar at z=5 (separation of 
21\arcsec, 136~kpc) for which \xmm\ archival data allow us to detect 
the two components separately. 
Using optical spectra we derived bolometric luminosities, BH masses 
and Eddington ratios that were compared to those of luminous SDSS quasars in 
the same redshift ranges.  
We find that the brighter component of both quasar pairs at $z\approx$~3.0--3.3 has high luminosities compared to the distribution of SDSS quasars at similar redshift, with J1622A having an order magnitude higher luminosity than the median. This source lies at the luminous end of the $z\approx$~3.3 quasar luminosity function. While we cannot conclusively state that the unusually high luminosities of our sources are related to their having a close companion, for J1622A there is only a 3\% probability that it is by chance. 

\end{abstract}




\section{Introduction}
Hierarchical merger models of galaxy formation predict that binary Active Galactic Nuclei (AGN) should be common in galaxies \citep{haehnelt02, volonteri03} over a limited time span. Understanding the types of galaxies and specific merger stages where AGN pairs preferentially occur may provide important clues on the peak black hole (BH) growth during the merging process \citep{begelman80, escala04}. Furthermore, multiple mergers offer a potential physical mechanism linking galaxy star formation with AGN feeding and BH-host galaxy coevolution (e.g., \citealt{silk98, dimatteo05, hopkins08}). 
Galaxy interactions are likely to produce strong star formation and to convey large amounts of gas into the nuclear galactic regions, thus feeding and obscuring the accreting BHs (which then become active). In a subsequent phase, the radiative feedback from the AGN can sweep the environment from the surrounding gas, making the SMBH shine as an unobscured AGN. Since the importance of galaxy mergers increases with redshift (e.g., \citealt{conselice03, lin08, lopez13, tasca14}), we may identify the high-redshift Universe as the ideal laboratory for finding binary AGN, thus witnessing the key early phase of quasar evolution. 

In the last decade, systematic studies, mostly based on the Sloan Digital Sky Survey (SDSS) at low ($z\simlt0.2$) redshifts, have provided significant numbers of BH binary AGN (e.g., \citealt{liu11, liu12}) and dual\footnote{In the following, with the term ``binary" we will refer to close ($\approx$ pc scale) systems; all the remaining, on $\approx$ kpc scales, will be referred to as dual.}  BH candidates by searching mostly among double-peaked \oiii\ AGN (e.g., \citealt{fu11, shen11a, comerford12, shangguan16}; see also \citealt{yuan16}), revealing a fraction as high as 3.6~per~cent of SDSS AGN pairs at low redshifts (after correcting for incompleteness; \citealt{liu11}). A fraction as high as 10~per~cent has been found by \cite{koss12} for the AGN selected at hard X-ray energies by the BAT camera onboard \swift. 
Indications for enhanced star-forming activity coupled to BH accretion in pairs are currently present (e.g., \citealt{ellison11, ellison13, green11, liu12, donley18}), especially at low (<40~kpc) separations \citep{ellison11}, but a physical coupling of accretion and star formation requires further and deeper investigation, including a complete knowledge of all selection effects. 
In this context, X-rays, because of their high penetrative power, provide an important probe of the active phase of AGN in pairs, and often represent a unique and ultimate tool in the hunt for multiple active nuclei in a galaxy, being less affected by contamination and absorption (e.g., \citealt{komossa03, ballo04, guainazzi05, bailon07, bianchi08, piconcelli10, mudd14, comerford15}). X-rays provide also a strong tool to pinpoint the early stage of interaction among the nuclei. 
Although obscured accretion is predicted by many models of galaxy mergers, it has not been found ubiquitously among these systems (e.g., \citealt{green10, green11}); this result can be, at least partially,  explained by the selection criteria adopted in the quest of dual and binary systems.  

Despite the many observational advances in discovering dual AGN and binary systems over the last decade (see e.g. the compilation summarized in Fig.~1 of \citealt{deane14} and the reviews by \citealt{bogdanovic15} and \citealt{komossa16}), many questions remain without a proper answer; among all, the link between AGN activation and merger rate, the activation radius (i.e., the typical separation at which the galactic nuclei become active during a merging process, see \citealt{mortlock99}), and the overall fraction of AGN in dual systems (hence, as a function of the separation). 
Furthermore, most of the available samples are currently limited to very low redshifts ($z\approx0.1-0.2$; e.g., \citealt{liu11, liu12, imanishi_saito14}; see also \citealt{ricci17}), leaving the high-redshift Universe an almost uncharted territory. 
A first, fundamental step in this direction requires a proper characterization of the known AGN pairs at high redshift. The present work, although unable of providing an answer to the issues described above because of the limited sample size and selection, is meant to start defining the properties of high-redshift dual quasars in terms of accretion and spectral energy distribution (i.e., corona/disc ratio, as provided by the \aox, the slope of an hypothetical power law connecting the UV to the soft X-rays). 
Furthermore, 
quasar pairs are expected to trace rich environments, especially at high redshift; however, in this regard, recent results suggest that quasar pairs are not always associated with significant overdensity of galaxies (see, e.g., the introduction in \citealt{onoue18}). Due to the lack of deep optical and near-IR imaging in the quasar fields presented in the following, we cannot investigate this issue in this paper. 

Here we present the multi-wavelength properties of three quasar pairs at $z$=3.02 and $z$=3.26 (selected from \citealt{hennawi10}, H10 hereafter), and $z$=5.02 \citep{mcgreer16}, focusing on their \xray\ emission thanks to proprietary \chandra\ and archival \xmm\ data. 
Given the assumed cosmology with $H_{0}$=70~km~s$^{-1}$~Mpc$^{-1}$, $\Omega_{\Lambda}$=0.73 and $\Omega_{M}$=0.27, 
the projected (proper transverse)  separations\footnote{One arcsec at $z=3.02$, $z=3.26$ and $z=5.02$ corresponds to $\approx7.9$, 7.5 and 6.5~kpc, respectively.} of the pairs is 43--65~kpc for the systems at $z\approx3-3.3$ and 136~kpc for the quasar pair at $z\approx5$, which are within  the spatial resolution capabilities of current \xray\ facilities at high redshift. 
To our knowledge, this work represents the first \xray\ investigation of quasar pairs at such high redshifts.

\section{Sample selection}
The three quasar pairs presented in this paper have been chosen via a twofold strategy: the first two 
pairs at $z\approx3$ were selected from the sample of H10, while the $z=5$ system was included in our analysis  after the discovery reported by \cite{mcgreer16}. Below we report on their original selection.  

Using color-selection and photometric redshift techniques, H10 searched more than 
8000~deg$^{2}$ of SDSS imaging data for dual quasar candidates at high redshift 
and confirmed them via follow-up spectroscopic observations. 
They found 27 high-redshift dual quasars ($z\approx2.9-4.3$), with proper 
transverse separations in the range 10--650~kpc. These dual quasars constitute 
rare coincidences of two extreme super-massive black holes (SMBHs), with masses 
above 10$^{9}$~\msun\ \citep{shen08}, which likely represent the highest peaks in the 
initial Gaussian density fluctuation distribution (e.g., \citealt{efstathiou88}). 
As such, these objects provide the opportunity for probing the hierarchical process of 
structure formation during the assembly of the most massive galaxies and SMBHs observable now. 
At present, this is the only sizable sample of high-redshift quasar pairs available, 
and represents an ideal test case to search for source over-densities. 
Taking into account the completeness of the H10 sample ($\approx$~50~per~cent), 
high-redshift quasar pairs are extremely rare, with a comoving number 
density of $\approx$~one dual per 10~Gpc$^{3}$ (at $z=3.5-4.5$), that is an order of magnitude lower 
than the extremely rare $z\approx6$ SDSS quasars (e.g., \citealt{fan01, jiang16}). 
In the H10 sample, eight pairs have a proper transverse separation 
below 100~physical kpc (angular separation $<$11\arcsec). 
Two of these eight quasar pairs were targeted by \chandra\ taking full advantage of its excellent on-axis spatial resolution and sensitivity to faint source detection and characterization; in particular, we observed SDSS~J1307$+$0422 (hereafter J1307) quasar pair at published redshifts $z=3.021$ and $z=3.028$ (A and B components, with a separation of 8.2\arcsec, corresponding to 65~kpc), and SDSS~J1622$+$0702 (J1622 hereafter) quasar pair at $z=3.264$ and $z=3.262$ (A and B components, with separations of 5.8\arcsec, i.e. 43~kpc). 
The difference in redshift in the A and B components of each quasar pair cannot be translated easily into peculiar velocities or distance along the line of sight between the two components; 
most likely, it reflects the systemic redshift uncertainties (which can be as large as 1000~km$^{-1}$; see Sect.~3.2 of H10) due to their classification as broad-line quasars.
We note that J1307B and J1622A show clear broad absorption lines (BAL) in the optical spectra (see Fig.~10 of H10); these absorption features are indicative of outflowing winds. The selection of these targets for \chandra\ follow-up observations was originally meant to search for possible indications of winds also in the \xray\ band; however, this kind of investigations typically requires a better photon statistics than what we achieved with our observations (unless very deep absorption features are present).

Both members of the last quasar pair (CFHTLS~J0221$-$0342, J0221 hereafter) presented in this paper have been identified as quasar candidates using color selection techniques applied to photometric catalogs from the Canada-France-Hawaii Telescope Legacy Survey (CFHTLS). Spectroscopic follow-up observations have shown that the redshift of the pair is $z=5.02$, with no discernible offset in redshift between the two objects. Their separation is 21\arcsec, i.e.136~kpc. The difference in their spectra and spectral energy distributions implies that they are not lensed images of the same quasar. The same consideration applies for the H10 quasar pairs discussed above.


\section{X-ray data reduction and analysis}
In the following we present the reduction and analysis of the proprietary \chandra\ data for J1307 and J1622, and of the \xmm\ archival data for J0221. Table~\ref{the_sample} reports the quasars analysed in this paper, including the exposure time, the number of net (i.e., background-subtracted) counts, and the signal-to-noise ratio (SNR).
\begin{table*}
\caption[]{The sample of quasar pairs: optical information and \xray\ data.}
\label{the_sample}
\begin{tabular}{cccccccc}
\hline
\noalign{\smallskip}
Src. & $z$ & RA & Dec & N$_{H,Gal}$ & Exp.Time & Net Counts & SNR \\
(1) & (2) & (3) & (4) & (5) & (6) & (7) & (8) \\
\hline
\multicolumn{8}{c}{\bf Chandra data} \\ \\
J1307A & 3.026$\pm{0.002}$ & 13:07:56.73 & +04:22:15.6 & 2.0 & 64.2 & 845 & 29.1\\
J1307B & 3.030$\pm{0.003}$ & 13:07:56.18 & +04:22:15.5 & 2.0 & 64.2 & 13 & 3.4 \\
J1622A & 3.264 & 16:22:10.11 & +07:02:15.3 & 4.5 & 65.1 & 215 & 14.7 \\
J1622B & 3.262 & 16:22:09.81 & +07:02:11.5 & 4.5 & 65.1 & 30 & 5.3 \\
\hline
\multicolumn{8}{c}{\bf XMM-Newton data} \\ \\
J0221A & 5.014$\pm{0.002}$  & 02:21:12.61 & -03:42:52.2 & 2.1 & 65.3/87.0/88.0 & 180/90 & 12.8/8.5\\
J0221B & 5.02  & 02:21:12.31 & -03:42:31.8 & 2.1 & 65.3/87.0/88.0 & 50/... & 6.1 \\
\hline
\end{tabular}
\vskip0.01cm
Notes --- The exposure times in the \xmm\ observation refer to pn/MOS1/MOS2, while the number of net counts and SNR to pn/MOS1+2 for J0221A and to the pn only for J0221B. 
(1) Source name as reported in the paper; 
(2) redshift with errors are derived from our own analysis (from the \cii\ 1334\AA\ line, when observed; see $\S$\ref{optical}); for the remaining objects, we report the published redshift of the source with no associated uncertainty. We note that the typical systemic redshift uncertainties can be as large as 1000~km$^{-1}$ (see Sect.~3.2 of H10), corresponding to $\Delta\ z>0.01$, since the redshift is measured from broad lines; (3) optical right ascension and (4) declination, both in J2000; (5) Galactic column density, in units of 10$^{20}$~cm$^{-2}$ (from \citealt{kalberla05}); (6) exposure time (in ks) used in our analysis (after removal of the periods of flares in the case of the \xmm\ observation); (7) source net (i.e., background-subtracted) counts in the $\approx$0.5--7~keV band; (8) source signal-to-noise ratio in the analysed \xray\ spectra.
\end{table*}


\begin{figure}
\includegraphics[width=\columnwidth,angle=0]{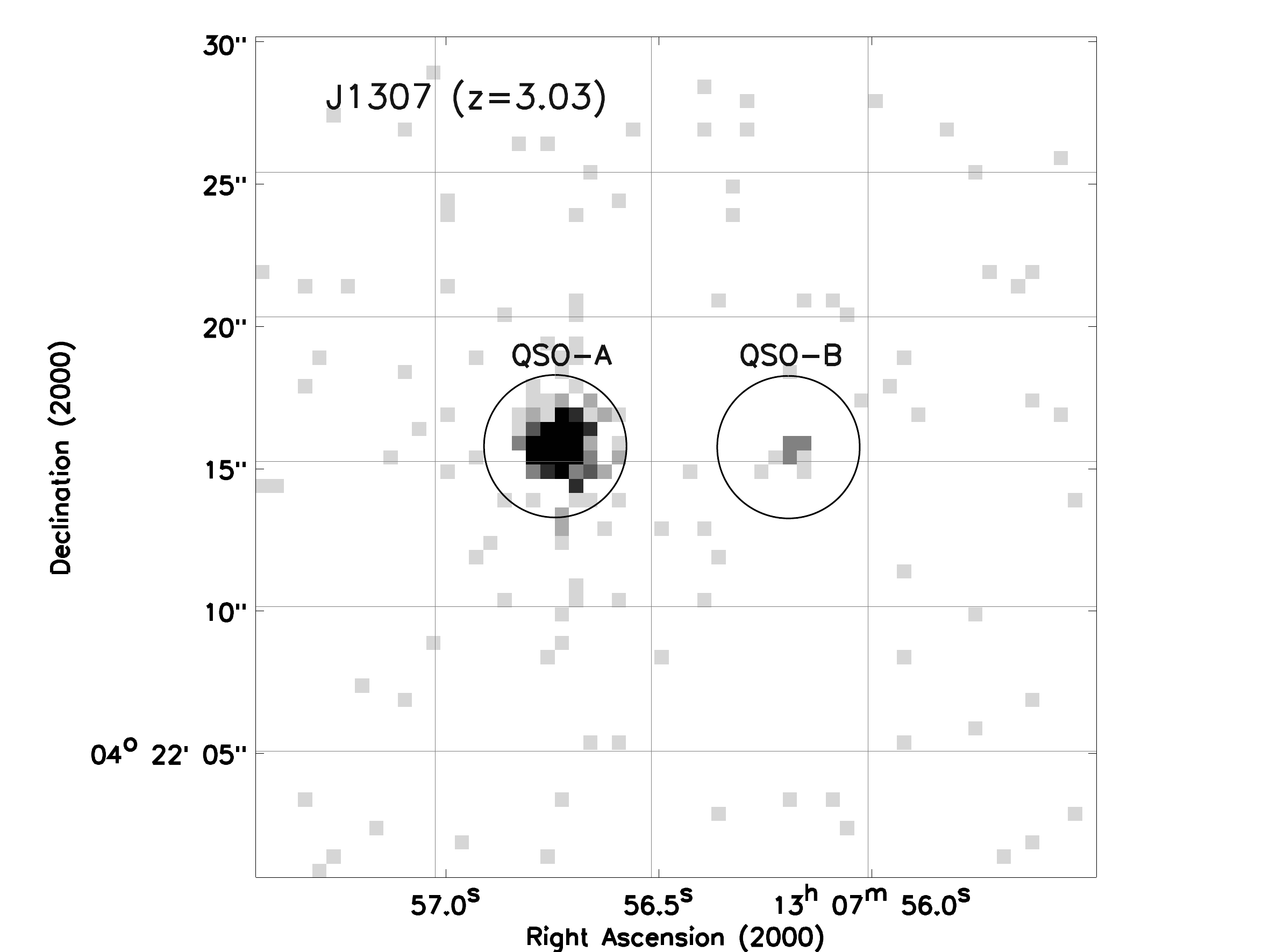}
\includegraphics[width=\columnwidth,angle=0]{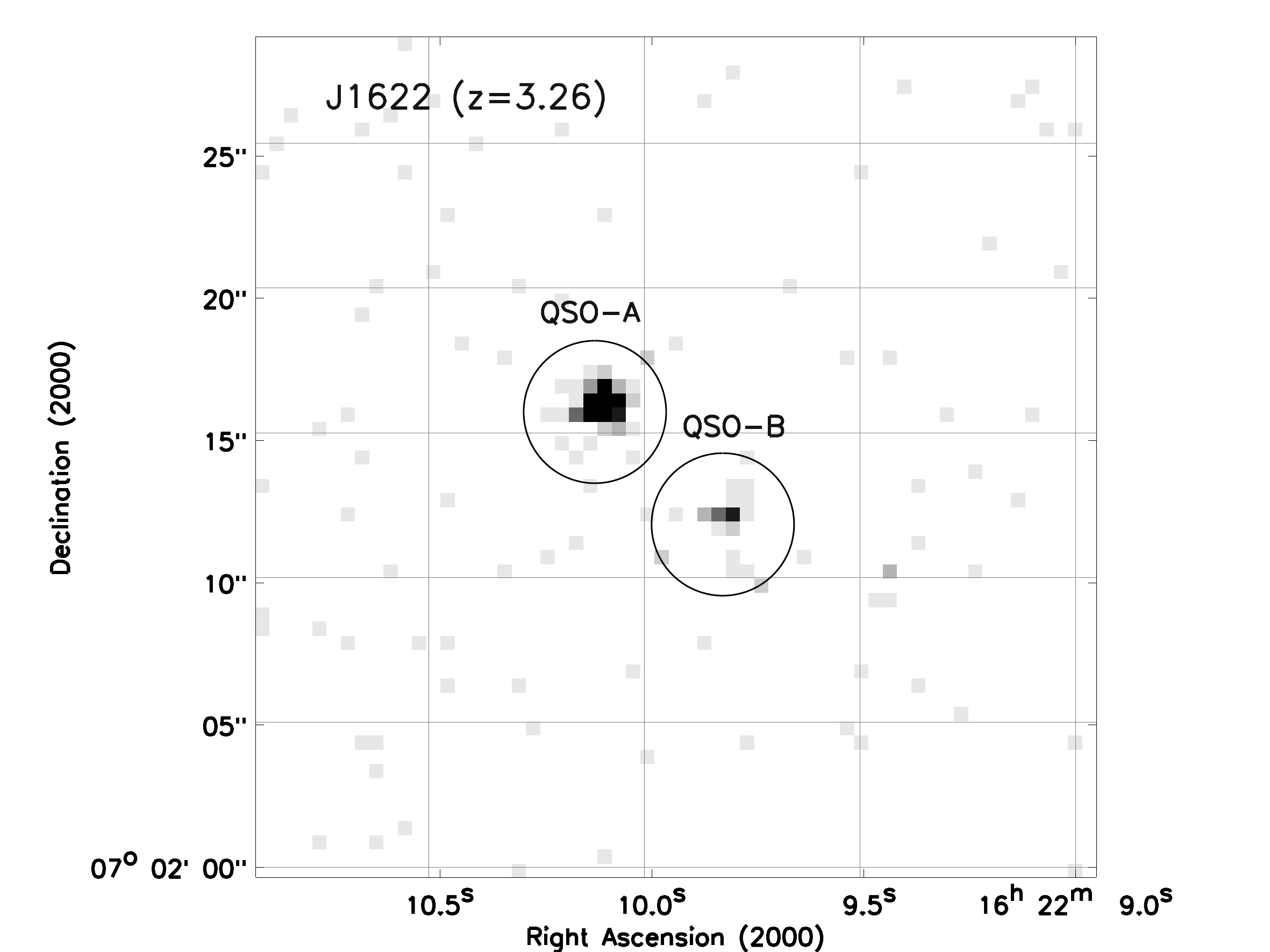}
\includegraphics[width=\columnwidth,angle=0]{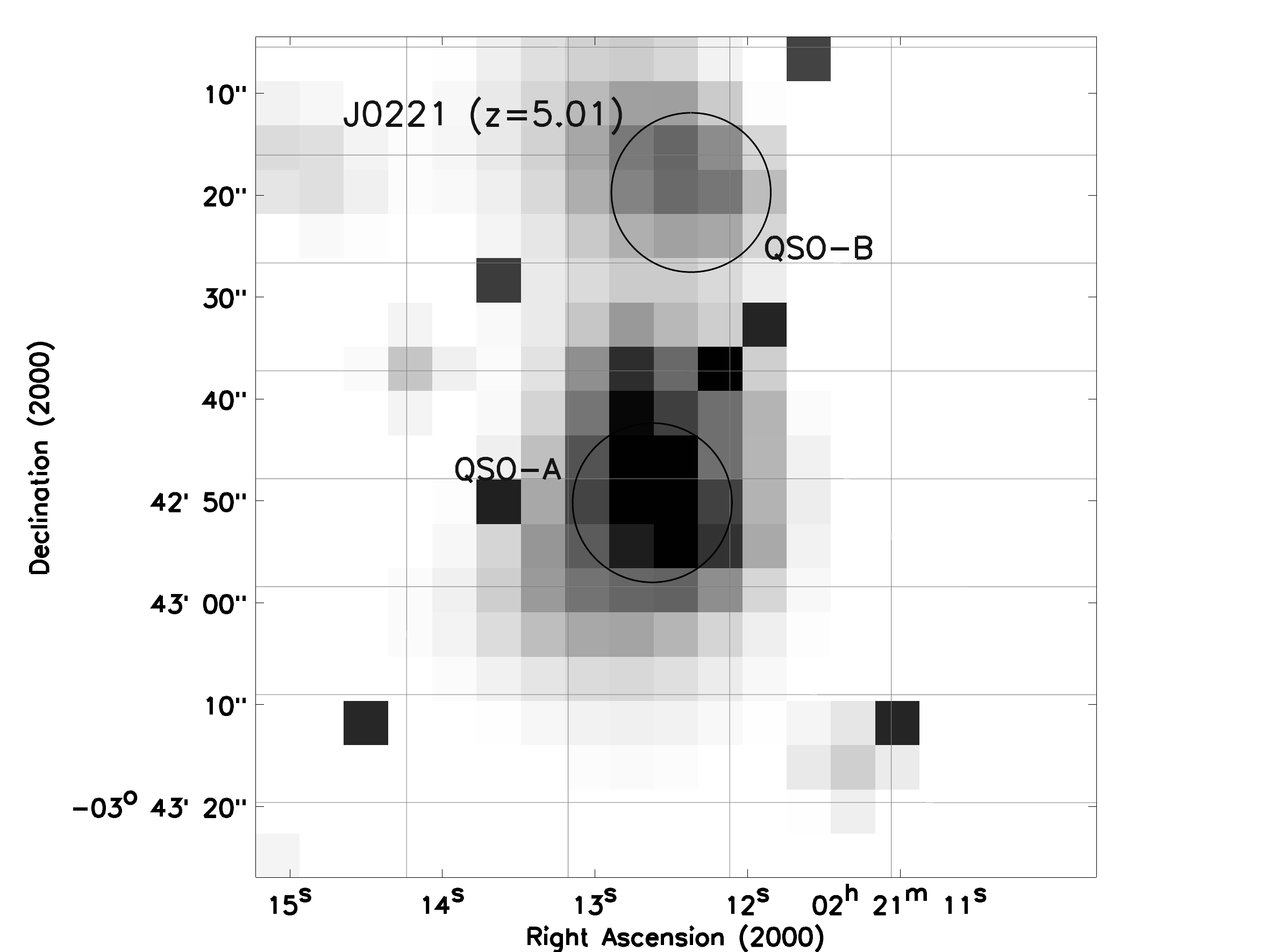}
\caption{0.5--7~keV band \chandra\ images of J1307 (top) and J1622 (middle) quasar fields. Each box side is 30\arcsec; the circles (diameter=5\arcsec) mark the quasars of the pairs. The QSO field of J0221 as observed by \xmm\ (pn camera, 0.3--3~keV band) is shown in the bottom panel; each side of the box is $\approx$85\arcsec, and circles have a diameter of 15\arcsec. All images keep the original pixel size 
($\approx$0.5\arcsec\ for \chandra\ and $\approx4.1$\arcsec\ for \xmm/pn).}
\label{xray_cutouts}
\end{figure}

\subsection{Chandra data reduction}
J1307 and J1622 were observed by \chandra\ in Cycle~15 on April 28$^{th}$  and June 1$^{st}$, 2014, with the ACIS-S3 CCD at the aimpoint, for an effective exposure time of 64.22~ks and 65.06~ks, respectively. 
Source spectra were extracted using the {\sc CIAO} software (v. 4.8) in the 0.5--7~keV band using a circular region with an extraction radius of 3.5\arcsec\ and 1.7\arcsec\ (J1307 A and B), and 2.5\arcsec\ for both J1622 quasars. In all cases the background spectra were extracted from larger regions close to the source, avoiding the  contribution from other sources. 
The number of source net counts are 845 (J1307A), 13 (J1307B), 215 (J1622A) and 30 (J1622B); spectra were therefore rebinned to 15 and 10 counts per bin for the two quasars with most counts, in order to apply the $\chi^{2}$ statistics, and to one count per bin for the sources with limited counting statistics; for these objects, the Cash statistics was used \citep{cash79}. 
All the spectral analyses were carried out using the {\sc xspec} package (v.12.8; \citealt{arnaud96}). 
\chandra\ cutouts in the 0.5--7~keV band are shown in the top (J1307) and middle (J1622) panels of Fig.~\ref{xray_cutouts}. 

\subsection{XMM-Newton data reduction}
J0221 was observed by \xmm\ three times, twice within the XMM-LSS project (PI: M. Pierre) with nominal exposures of $\approx14-15$~ks, and once to observe the low-mass cluster XLSSC~006 (PI: F. Pacaud) for a nominal exposure of $\approx102$~ks. In the following we report the analysis of the data having the longest exposure. The data were reprocessed using the {\sc SAS} software (v.15); high flaring background periods were removed with a sigma-clipping method, leaving a final exposure time of 68.3, 87.0 and 88.0~ks for the pn, MOS1 and MOS2 cameras, respectively. Source spectra for both quasar components were extracted from a circular region centered on their optical position; we used a radius of 15\arcsec\ (corresponding to about 70~per~cent of the pn encircled energy fraction at the source position) to maximise the source SNR and to prevent significant contamination by the relatively bright A component towards the much fainter B component; background spectra were extracted from a 30\arcsec\ circular region in the same CCD as the source. 
We also checked that the apparently extended \xray\ emission around J0221A (see the bottom panel of Fig.~\ref{xray_cutouts}) was due to the relatively broad pn Point Spread Function (PSF); for this check, we used the SAS routine {\sc eradial}, which allows a comparison between the source count distribution and the nominal PSF at a given position, fitted to the actual radial profile data.  
At the end of the spectral extraction procedure, the number of source net counts in the 0.5--7~keV band is 
180 (pn) and 90 (MOS1$+$2) for the A component of the pair, and 50 for the fainter B component (only pn data were extracted). 
MOS1 and MOS2 spectra for J0221A were summed and, similarly to pn data, were grouped to have at least 10 counts per bin to apply the $\chi^{2}$ statistics; for the much fainter B component, Cash statistics was adopted in fitting the spectrum; a binning of one count per bin was applied in this case. 
The pn 0.3--3~keV image (maximising the emission of the faint B component) of J0221 is shown in the bottom panel of Fig.~\ref{xray_cutouts}.

\begin{figure*}
\centering
\includegraphics[width=0.5\columnwidth,angle=-90]{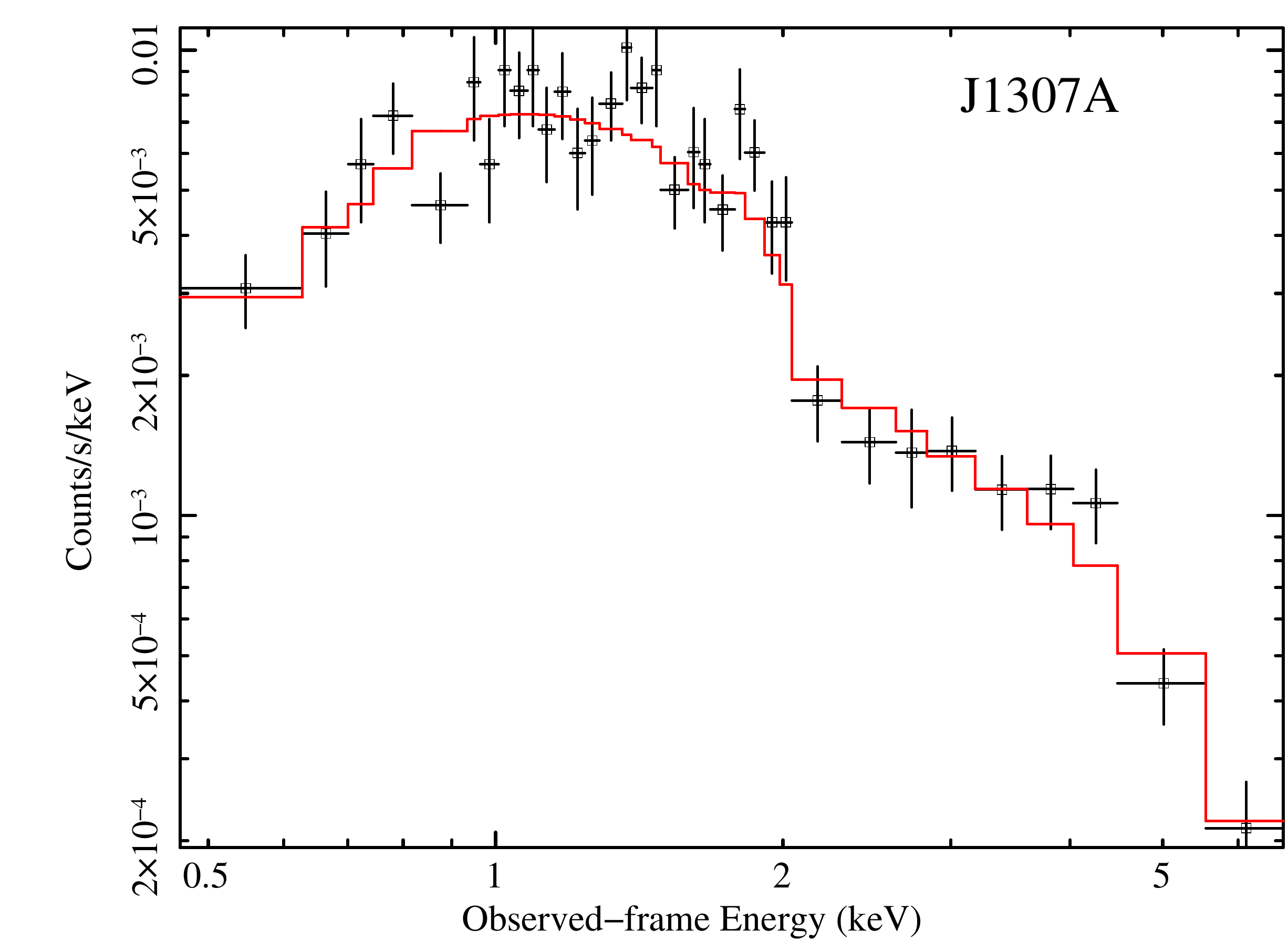}\hskip0.2cm
\includegraphics[width=0.5\columnwidth,angle=-90]{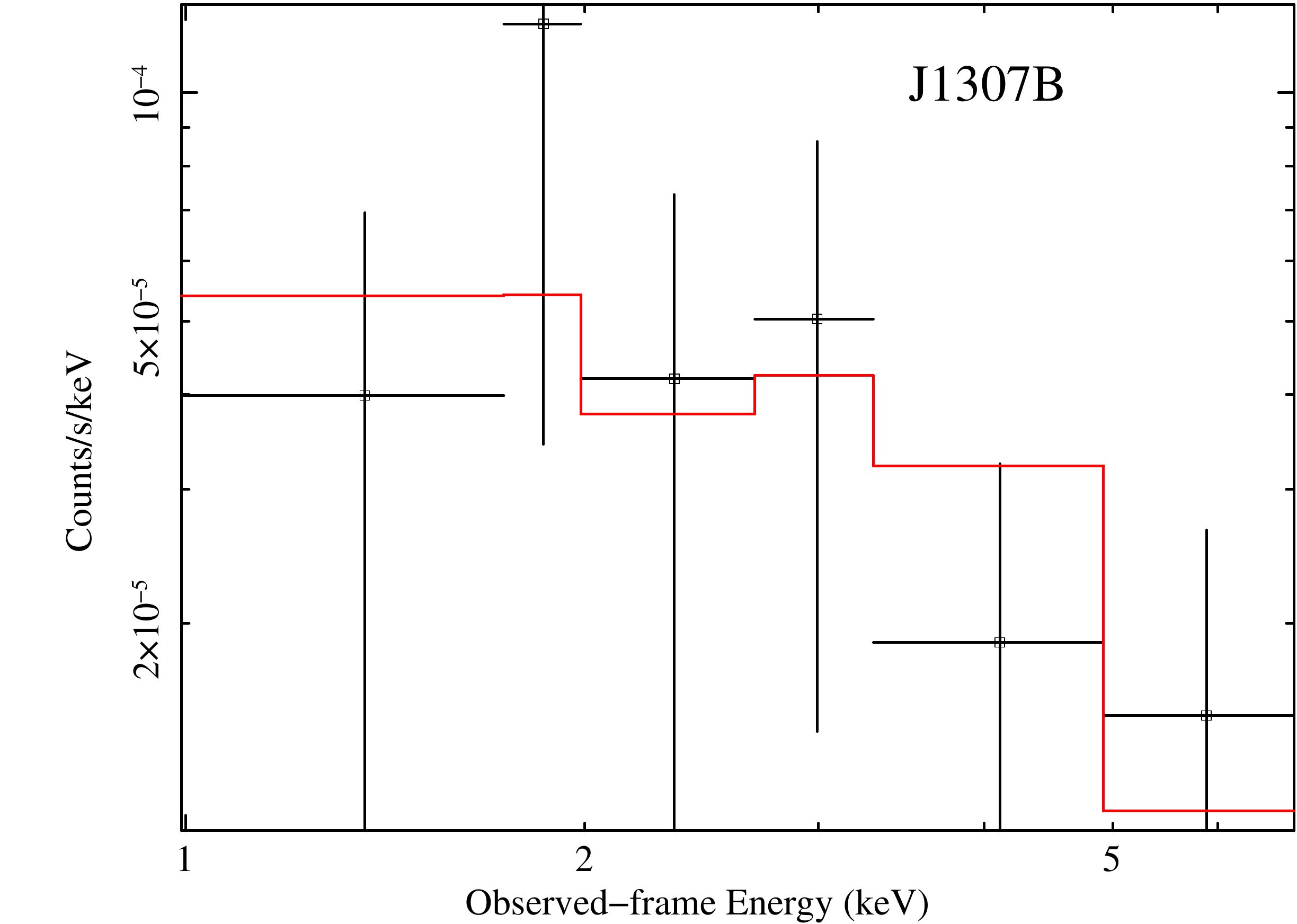}\vspace{0.5cm} 
\includegraphics[width=0.5\columnwidth,angle=-90]{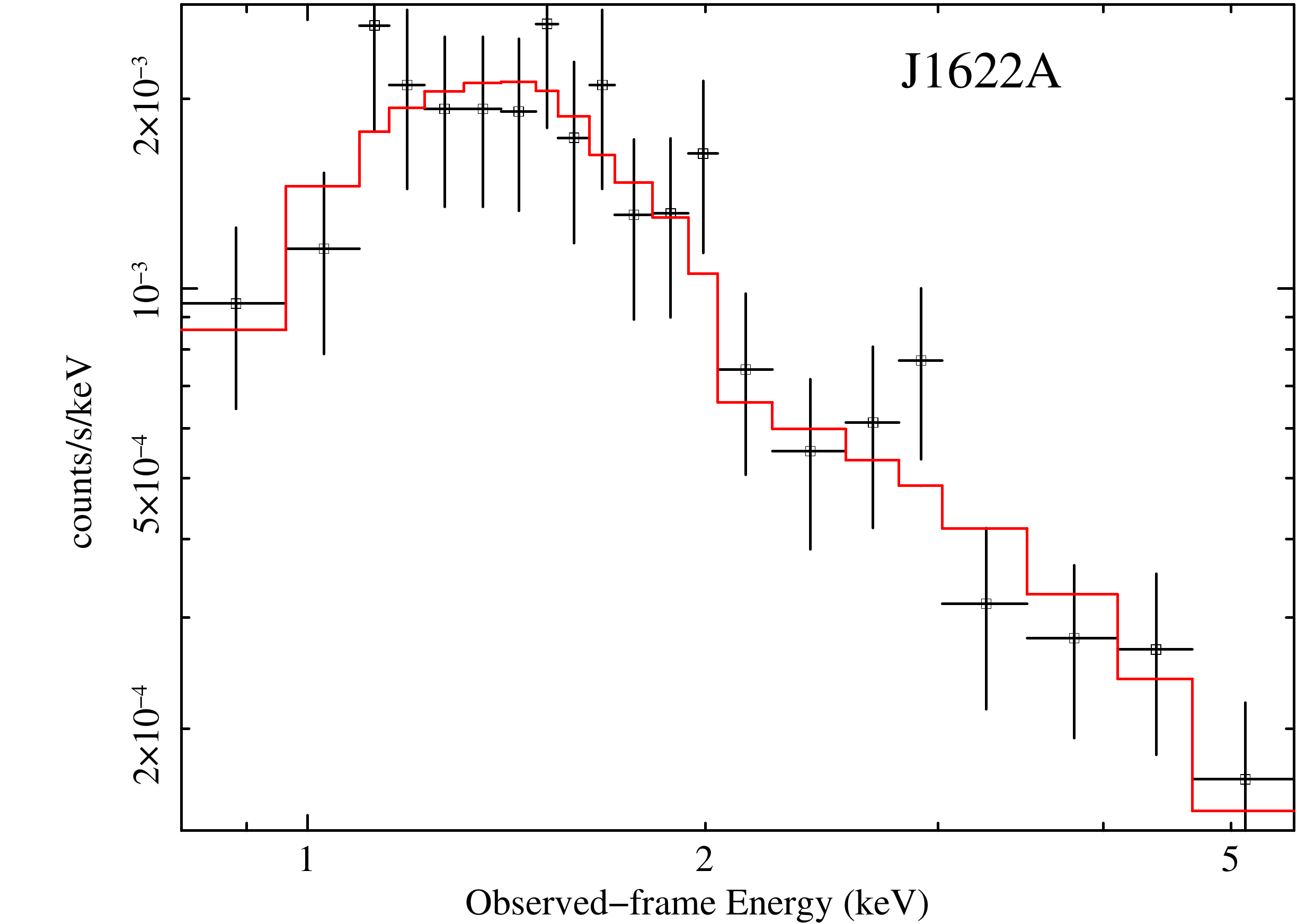}\hskip0.2cm
\includegraphics[width=0.52\columnwidth,angle=-90]{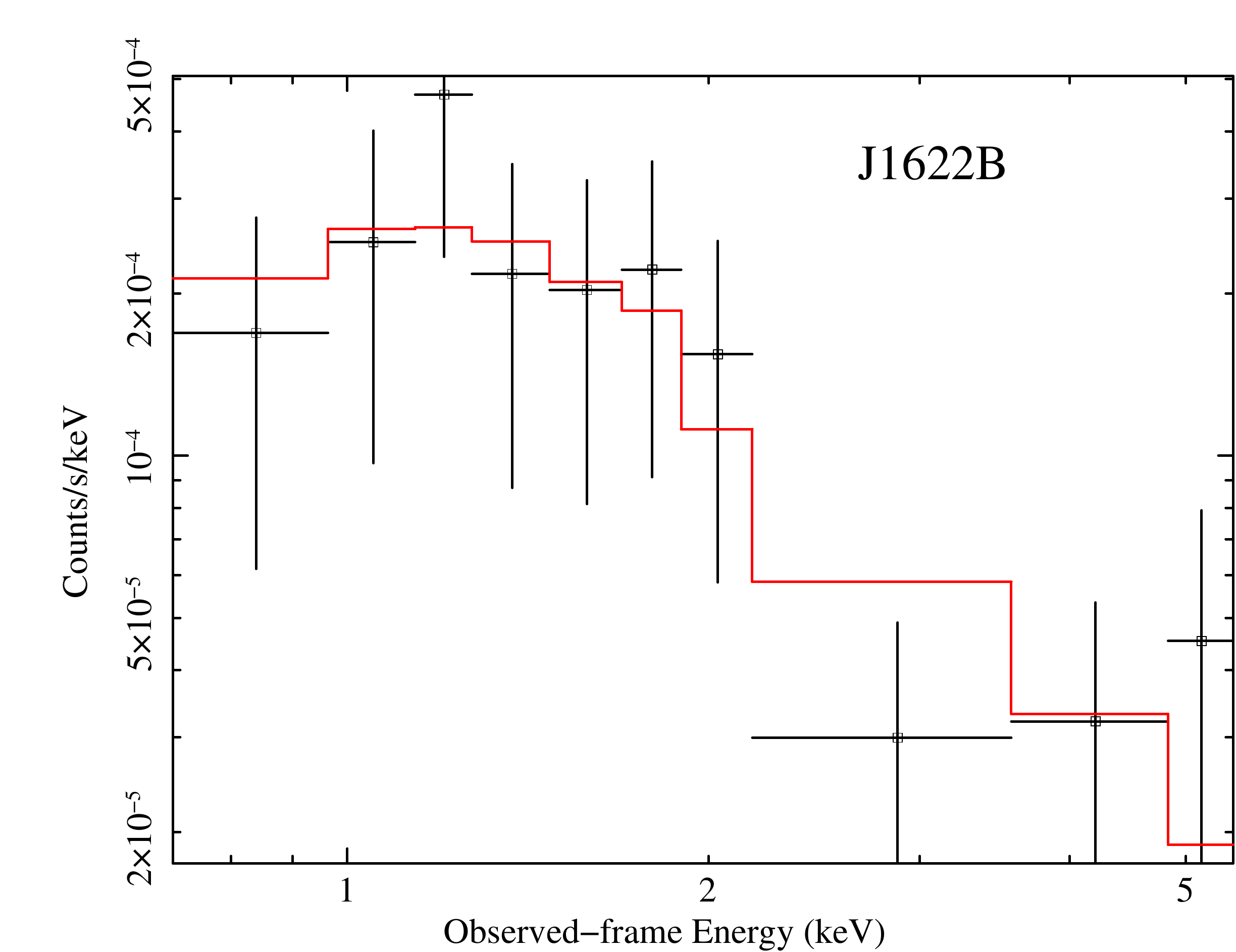}
\includegraphics[width=0.5\columnwidth,angle=-90]{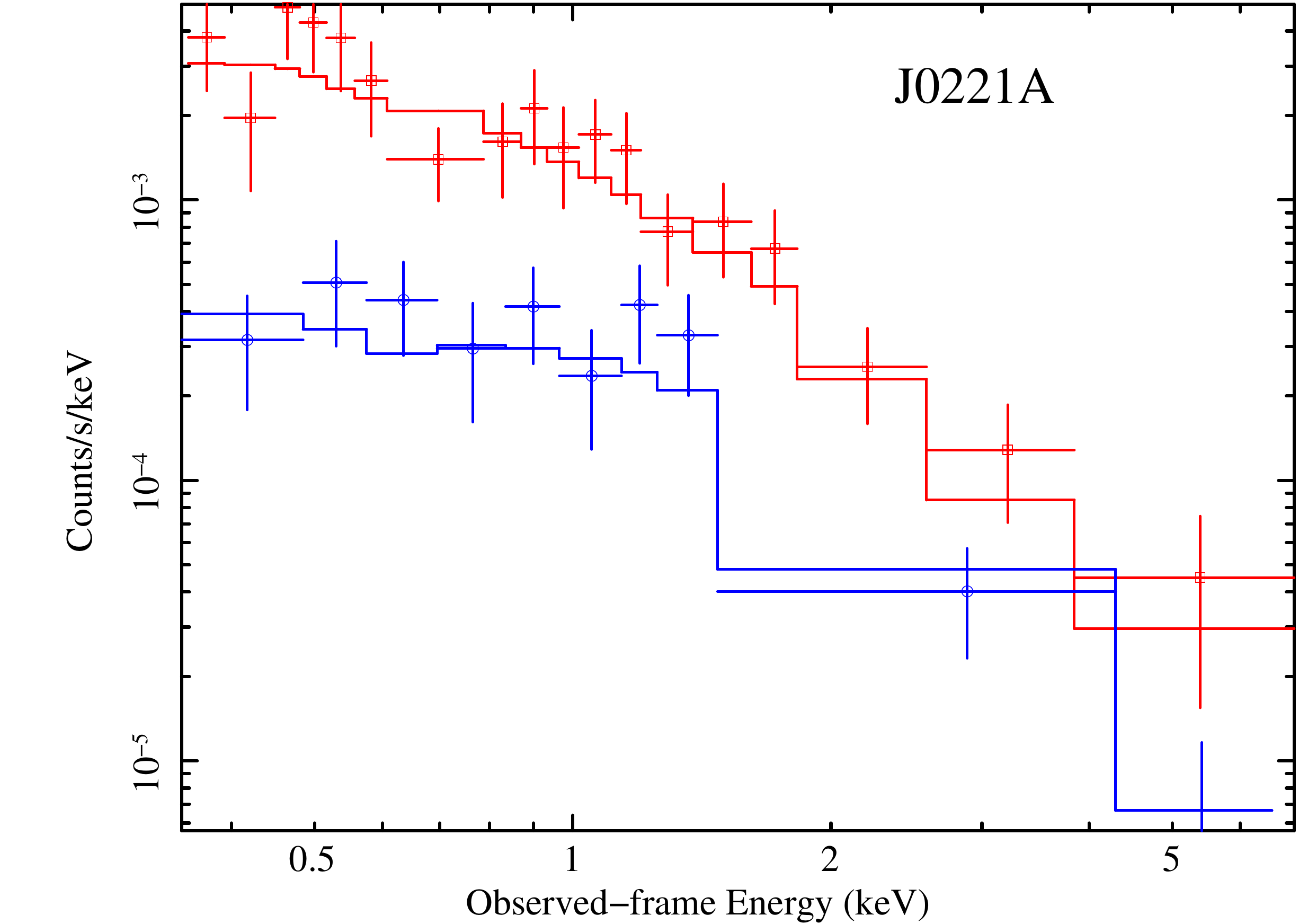}\hskip0.2cm
\includegraphics[width=0.5\columnwidth,angle=-90]{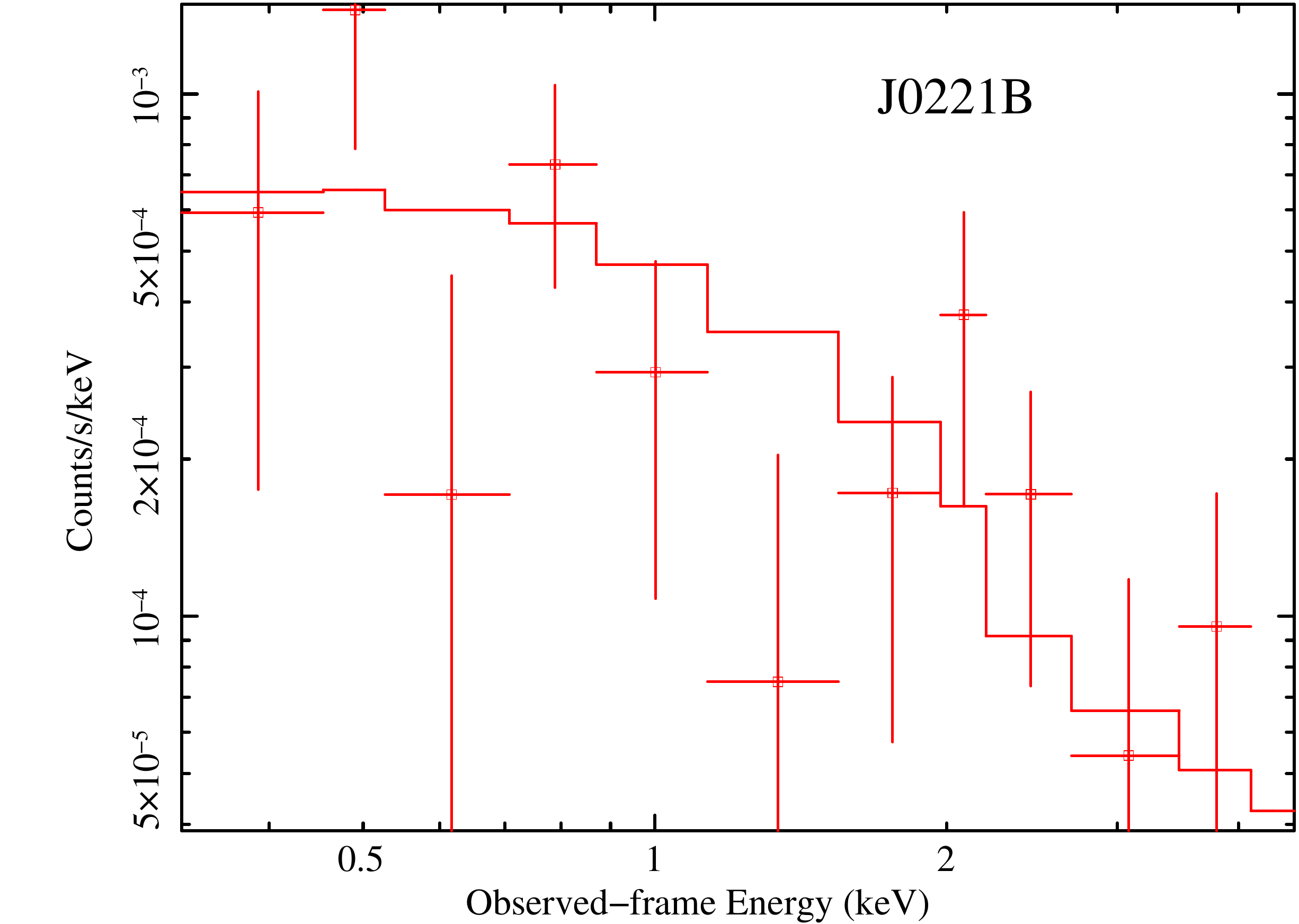}
\caption{\chandra\ (top two rows) and \xmm\ (bottom row) spectra of the three quasar pairs with best-fitting  spectrum (continuous curve). The strong rebinning is for presentation purposes. For J0221A, both pn (red) and combined MOS1+2 (blue) spectra are shown; for J0221B, only the pn spectrum was extracted.}
\label{xray_spectra}
\end{figure*}

\begin{table*}
\small
\caption[]{\xray\ spectral parameters of the quasar pairs derived from \chandra\ and \xmm\ analysis.}
\label{xray_results}
\begin{tabular}{cccccc}
\hline
\noalign{\smallskip}
\small
Src. & $\Gamma$ & N$_{\rm H}$ & F$_{2-10~keV}$ & L$_{2-10~keV}$ & $\chi^{2}$(C-stat)/d.o.f.\\
        &                   & (cm$^{-2}$)  & (\cgs)                  & (\lum)                  &                               \\
\hline
\noalign{\smallskip}
J1307A & 1.53$\pm{0.10}$ & & & & 44.3/48 \\
   & 1.8$\dagger$ & 1.88$^{+1.28}_{-1.02}\times10^{22}$ & 8.6$\times10^{-14}$ & 5.2$\times10^{45}$ & 52.0/48 \\
\hline
J1307B & 0.4$\pm{1.0}$ & & & & 9.7/10 \\
   & 1.8$\dagger$ & 7.2$^{+6.1}_{-4.7}\times10^{23}$ & 4.0$\times10^{-15}$ & 3.0$\times10^{44}$ & 8.3/10 \\
\hline
J1622A & 1.21$\pm{0.20}$ & & & & 15.5/19 \\
   & 1.86$^{+0.48}_{-0.44}$ & 1.77$^{+1.42}_{-1.13}\times10^{23}$ & 3.4$\times10^{-14}$ &  2.6$\times10^{45}$ & 8.2/18 \\
\hline
J1622B & 1.5$\pm{0.6}$ & & 4.5$\times10^{-15}$ & 1.9$\times10^{44}$  & 10.7/27 \\
   & 1.8$\dagger$ & $<2.1\times10^{23}$ & & & 10.6/27 \\
\hline
J0221A & $2.14^{+0.26}_{-0.24}$ & & 6.0$\times10^{-15}$ & 1.9$\times10^{45}$ & 34.6/38 \\
   & $2.22^{+0.50}_{-0.28}$ & $<4.5\times10^{22}$ & & & 34.5/37 \\
\hline
J0221B & $1.6^{+0.6}_{-0.5}$ & & 5.3$\times10^{-15}$ & 6.9$\times10^{44}$  & 91/114 \\
   & 1.8$\dagger$ & $<4.7\times10^{22}$ & & & 85.3/114 \\
\noalign{\smallskip}
\hline
\end{tabular}
\vskip0.01cm
Notes --- $\dagger$ Fixed photon index. 
Fluxes are reported in the observed 2--10~keV band, while luminosities are intrinsic (i.e., corrected for obscuration, if present) and in the rest-frame 2--10~keV band. They are reported for the spectral fits which are considered to provide the best-fit description of the source emission. The last column indicates the quality of the fit in terms of either $\chi^{2}$ (A components) or C-stat (B components) over the number of degrees of freedom, depending on the statistics adopted in the spectral fitting (see text for details). 
\end{table*}


\subsection{X-ray spectral analysis}
Given the limited \xray\ photon statistics available for the quasars under investigation, we adopted  simple models, namely a power law and an absorbed power law. No additional component (e.g., reflection) seems to be present. All of the models take into account the Galactic absorption (\citealt{kalberla05}, see Table~\ref{the_sample}). In the following, errors and upper limits are reported at the 90~per~cent confidence level for one parameter of interest \citep{avni76}, unless stated otherwise. Upper limits to the equivalent width (EW) of the iron K$\alpha$ emission line at 6.4~keV are reported in the source rest frame. Errors on the \xray\ fluxes and luminosities are of the order of 6--30~per cent for the quasars with most counts, and 50--70~per cent for the quasars with poorer  statistics. 
\vskip0.2cm\pn
{\bf J1307A.}
A power law well reproduces the emission for the A component ($\chi^{2}$/dof, degrees of freedom = 44.3/48); a photon index of $\Gamma=1.53\pm{0.10}$ is obtained. 
If we ascribe the apparently flat photon index 
(wrt. $\Gamma\approx1.8-2.0$  typically observed in unobscured quasars; \citealt{piconcelli05}) to the presence of obscuration (i.e., fixing $\Gamma=1.8$ and leaving the column density free to vary), we derive \nh=1.88$^{+1.28}_{-1.02}\times10^{22}$~cm$^{-2}$, i.e., moderate obscuration is possibly present in J1307A  ($\chi^{2}$/dof=52.0/48). 
No iron emission line is required, the limit to the EW of the 6.4~keV component is 140~eV. The 2--10~keV flux is $\approx8.6\times10^{-14}$~\cgs, corresponding to an intrinsic (i.e., corrected for the  obscuration) rest-frame 2--10~keV luminosity of $\approx5.2\times10^{45}$~\lum. The best-fitting spectrum is shown in Fig.~\ref{xray_spectra} (top-left panel).

This source is the only quasar of the original H10 sample with a detection in the FIRST survey at 1.4~GHz \citep{becker95}; the flux density is $\approx$14.3~mJy. Using the available SDSS photometry and the definition of radio loudness as reported by Kellermann et al. (1989), 
\hbox{$R$=$f_{\rm 5~GHz}/f_{\rm 4400~\mbox{\scriptsize\AA}}$} (rest frame),\footnote{The rest-frame 5~GHz flux density is computed from the observed 1.4~GHz flux density assuming a radio power-law slope of $\alpha=-0.8$ (i.e., S$_{\nu}\propto \nu^{\alpha}$) while, to derive the rest-frame 4400\AA\ flux density, we used SDSS photometry following the procedure described in \cite{vignali03}.} we are able to define J1307A as moderately radio loud (R$\approx$30). 
We checked for the presence of \xray\ extension in this object, possibly related to some jet emission, by comparing the source count distribution vs. the PSF; we found no clear evidence for extension. 
However, the relatively flat photon index observed in this quasar ($\Gamma\approx1.5$) may be due to the presence of an unresolved jet and associated \xray\ emission (see \citealt{miller11}). \\ \\
{\bf J1307B.} 
This source is characterized by a \civ\ 1549\AA\ BAL feature in the SDSS spectrum, indicative of extinction in an outflowing wind. BALQSOs are often characterized by obscuration also in the \xray\ band (e.g., \citealt{green96,green01,shemmer05,giustini08}; see also \citealt{luo13} for a different scenario). This is likely the most viable explanation for J1307B, whose spectrum is parametrized by a flat photon index ($\Gamma=0.4\pm{1.0}$). The flat \xray\ slope supports the presence of obscuration; if this component is included in the spectral fitting and $\Gamma=1.8$ (fixed) is adopted, we obtain \nh=7.2$^{+6.1}_{-4.7}\times10^{23}$~cm$^{-2}$ and a good fit (C-stat/dof=8.3/10); the spectrum is shown  
in Fig.~\ref{xray_spectra}, top-right panel. 
The source 2--10~keV flux and intrinsic rest-frame 2--10~keV 
luminosity are $\approx4.0\times10^{-15}$~\cgs and $\approx3.0\times10^{44}$~\lum, respectively. 
The upper limit to the neutral iron K$\alpha$ EW is $\approx$2~keV. \\ \\
{\bf J1622A.}
Fitting the spectrum of J1622A with a power law results in $\Gamma=1.21\pm{0.20}$ ($\chi^{2}$/dof=15.5/19); the addition of absorption provides an improvement in the fitting ($\chi^{2}$/dof=8.2/18), and a more typical photon index of  $\Gamma=1.86^{+0.48}_{-0.44}$ is derived. The column density is \nh=1.77$^{+1.42}_{-1.13}\times10^{23}$~cm$^{-2}$. 
As for J1307B, J1622A can be optically classified as a BALQSO and is heavily obscured in X-rays. No iron line is detected (EW<300~eV). 
The source 2--10~keV flux (luminosity, corrected for the obscuration) is $\approx3.4\times10^{-14}$~\cgs\ ($\approx2.6\times10^{45}$~\lum). The best-fitting spectrum is shown in Fig.~\ref{xray_spectra} (middle-left panel). \\ \\
{\bf J1622B.} 
A power law with $\Gamma=1.5\pm{0.6}$ reproduces the \xray\ emission of J1622B reasonably well 
(C-stat/dof=10.7/27). The inclusion of obscuration at the source redshift provides \nh$<2.1\times10^{23}$~cm$^{-2}$ (assuming $\Gamma$=1.8). No iron line is detected, the EW upper limit being $\approx$~1.4~keV. The source 2--10~keV flux (luminosity) is $\approx4.5\times10^{-15}$~\cgs\ ($\approx1.9\times10^{44}$~\lum). The best-fitting spectrum is shown in Fig.~\ref{xray_spectra} (middle-right panel). \\

Summarizing, the analysis of the \chandra\ data indicates that the two quasars of the pairs which are classified as BALQSOs in the optical band are actually obscured in X-rays by column densities of the order of a few$\times10^{23}$~cm$^{-2}$. All quasar luminosities are above 10$^{44}$~\lum, with the two optically brightest members of the two pairs (the A quasars) having $L_{2-10~keV}>10^{45}$~\lum. No indication of further components (e.g., reflection, iron emission line) is apparently present in the spectra. All of the spectral results are reported in Table~\ref{xray_results}. \\ \\
\pn
{\bf J0221A.}
The pn and MOS1+2 spectra of J0221A were fitted simultaneously, leaving the normalizations of the pn and  combined MOS cameras free to vary to account for possible (though minor) inter-calibration issues. 
A power law model with $\Gamma=2.14^{+0.26}_{-0.24}$ is able to reproduce the \xray\ spectra well ($\chi^{2}$/dof=34.6/38), with no indication of further spectral complexities. 
The inclusion of obscuration at the source redshift provides a column density upper limit of $\approx4.5\times10^{22}$~cm$^{-2}$; the upper limit to the 6.4~keV iron line is $\approx$730~eV. 
The derived 2--10~keV flux (luminosity) is $\approx6\times10^{-15}$~\cgs\ ($\approx1.9\times10^{45}$~\lum), with uncertainties of the order of 40~per cent. The best-fitting spectrum is shown in Fig.~\ref{xray_spectra} (bottom-left panel). \\ \\
{\bf J0221B.}
The photon statistics for J0221B (pn data) allows a basic parameterization of its spectrum; a single power law provides a good fit the the data (C-stat=91/114) and returns $\Gamma=1.6^{+0.6}_{-0.5}$. The obscuration, if present, is below $\approx4.7\times10^{22}$~cm$^{-2}$ (with $\Gamma$=1.8); the upper limit to the line EW is $\approx$450~eV. The derived 2--10~keV flux and rest-frame luminosity are $\approx5.3\times10^{-15}$~\cgs\  and  $\approx6.9\times10^{44}$~\lum, respectively, with uncertainties of the order of 70~per cent. The best-fitting spectrum is shown in Fig.~\ref{xray_spectra} (bottom-right panel). \\ \\

\begin{table*}
\caption[]{Source parameters derived from the analysis of the optical spectra.}
\label{optical_results}
\begin{tabular}{ccccccccc}
\hline
\noalign{\smallskip}
Src. & FWHM & F$_{1350\AA}$ & Log(M$_{\rm BH}$/\msun) & L$_{\rm bol}$ & $\Delta$v & $\lambda_{\rm Edd}$ & Notes & SNR \\
(1) & (2) & (3) & (4) & (5) & (6) & (7) & (8) & (9) \\
\hline
\noalign{\smallskip}
J1307A & $7900\pm{170}$ &   $65\pm{2}$     & 9.5--10.0 & 27.8 & $-1800\pm{970}$    & 0.21--0.76 & SDSS BOSS                 & 8 \\	
J1307B & $5440\pm{225}$ &   $14.6\pm{0.8}$  & 9.0--9.3   &	  6.3 & $-2240\pm{1030}$  & 0.24--0.45 & SDSS BOSS$\dagger$ & 13 \\
J1622A & $4870\pm{40}$ & $158\pm{3}$     & 9.6--9.8   &	81.4 & $-1000$                  & 0.99--1.65 & Keck (H10)$\dagger$    & 40 \\	
J1622B & $4970\pm{480}$ &    $6.5\pm{1.5}$  & 8.7--9.1    &	  3.3 & $-1000$                  & 0.20--0.52 & SDSS BOSS                  & 4 \\ 
J0221A & $3050\pm{55}$ &  $13.1\pm{1.6}$  & 8.6--9.1    &	18.8 & $-2150\pm{610}$    & 1.07--3.30 &  SDSS                           & 5 \\ 
J0221B & $3810\pm{350}$ &   $2.2\pm{0.3}$  & 8.8            &	  3.2 &                                 & 0.40          & MMT$\ddagger$            & 3 \\
\noalign{\smallskip}
\hline
\end{tabular}
\vskip0.01cm
(1) Source name; (2) FWHM of the \civ\ line (km/s); (3) value of the source continuum at 1350\AA\ (in units of $10^{-17}$~\cgs\AA$^{-1}$), derived from the fitting of the optical spectrum; (4) logarithm of black hole mass (range) derived as described in $\S$\ref{optical}; the upper boundary is usually obtained  using Eq. (2) of \cite{shen11b}, the \civ\ FWHM and the 1350\AA\ continuum luminosity; (5) bolometric luminosity (in units of 10$^{46}$~\lum) derived from the 1350\AA\ luminosity and adopting a bolometric correction of 3.8, valid for luminous SDSS quasars (see \citealt{richards06}); (6) blueshift of the \civ\ line computed with respect to \cii\ (measured in km~s$^{-1}$). Reported 
errors have been estimated from Monte Carlo trials of mock spectra; see $\S\ref{optical}$ for details about J1622 quasars; 
(7) Eddington ratio range (where the bolometric luminosity is kept constant, while the Eddington luminosity takes into account the range of derived black hole masses); the Eddington ratio is defined as $\lambda_{\rm Edd}=L_{\rm bol}/L_{\rm Edd}$, where L$_{\rm Edd}$ is the Eddington luminosity ($\approx1.3\times10^{38}~(M_\mathrm{BH}/M_\odot$)~\lum); 
(8) origin of the optical spectrum used to derive the 1350\AA\ continuum flux and the mass of the BH;  (9) SNR of the optical continuum, close to the \civ\ line used to estimate the black hole mass. 
$\dagger$: classification as a BALQSO; $\ddagger$: provided by I. McGreer \citep{mcgreer16}.
\end{table*}


\section{Optical data analysis} 
\label{optical}
Archival observed-frame optical spectra of the QSO pairs have been analysed to estimate SMBH properties (black hole masses, bolometric luminosities and accretion rates).  

\begin{figure*}
\centering
\includegraphics[width=8.8cm,trim=60 55 0 0,clip,angle=180]{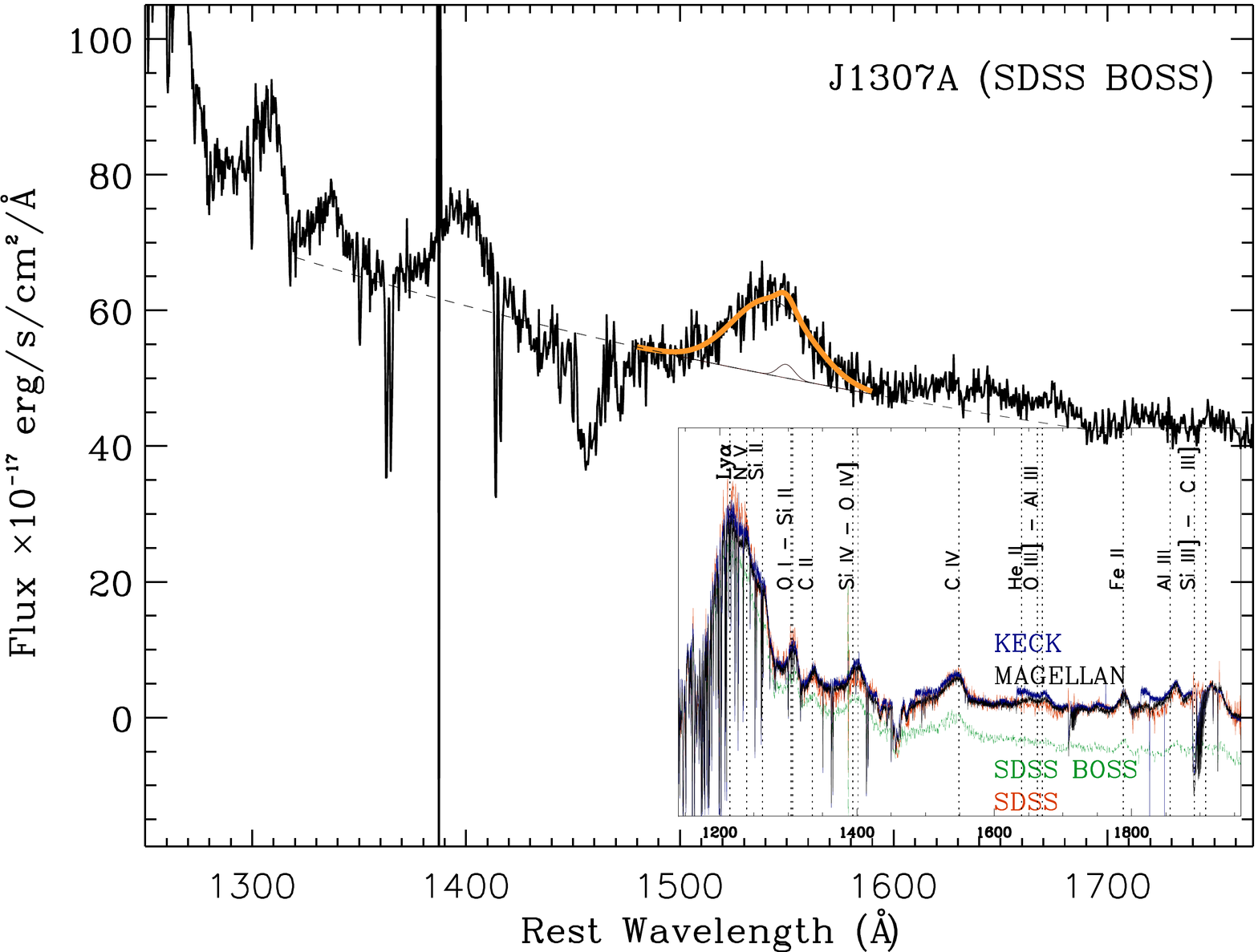}
\includegraphics[width=8.8cm,trim=60 55 0 0,clip,angle=180]{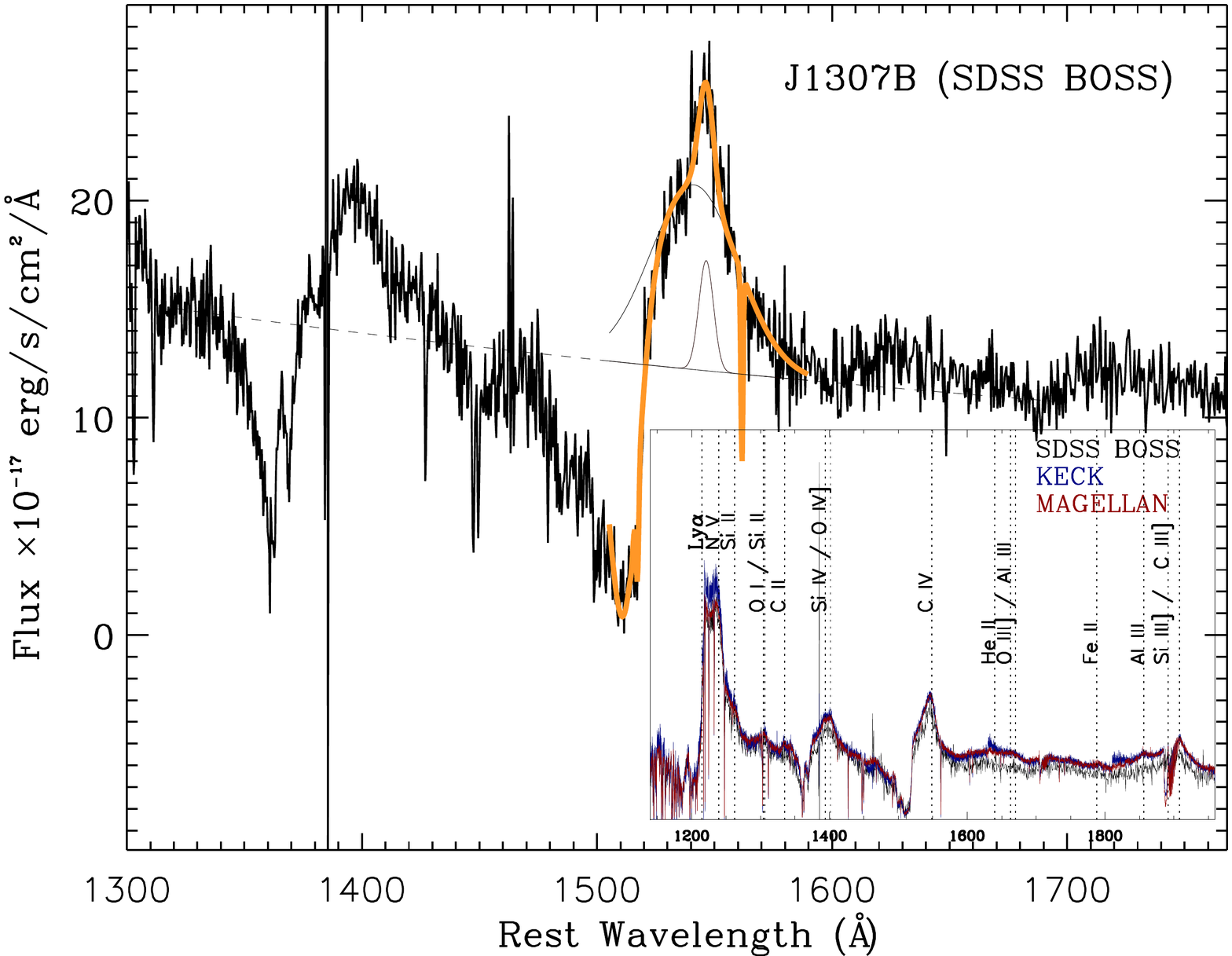}
\vskip0.4cm
\includegraphics[width=8.8cm,trim=35 55 0 0,clip,angle=180]{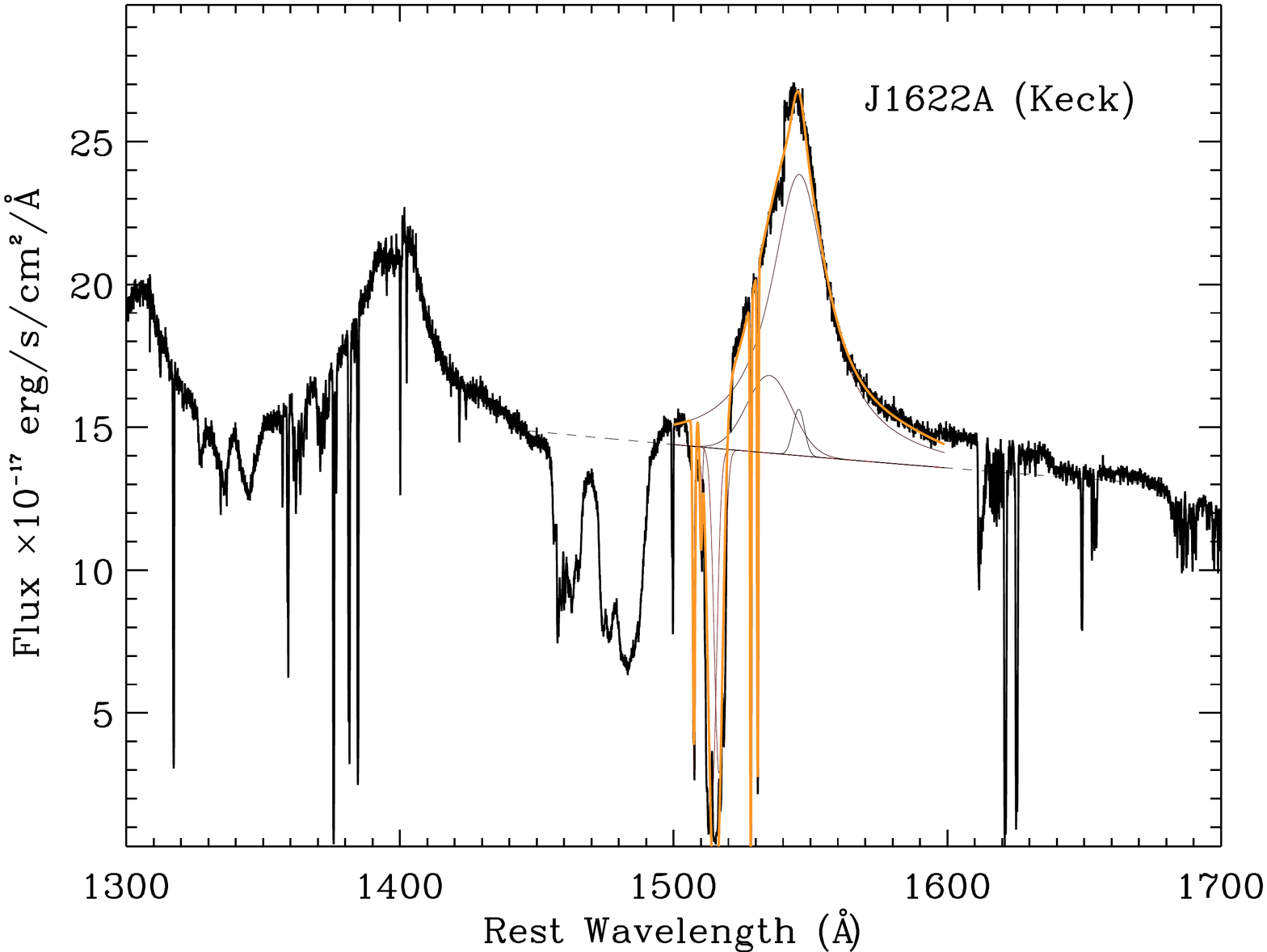}
\includegraphics[width=8.8cm,trim=60 55 0 0,clip,angle=180]{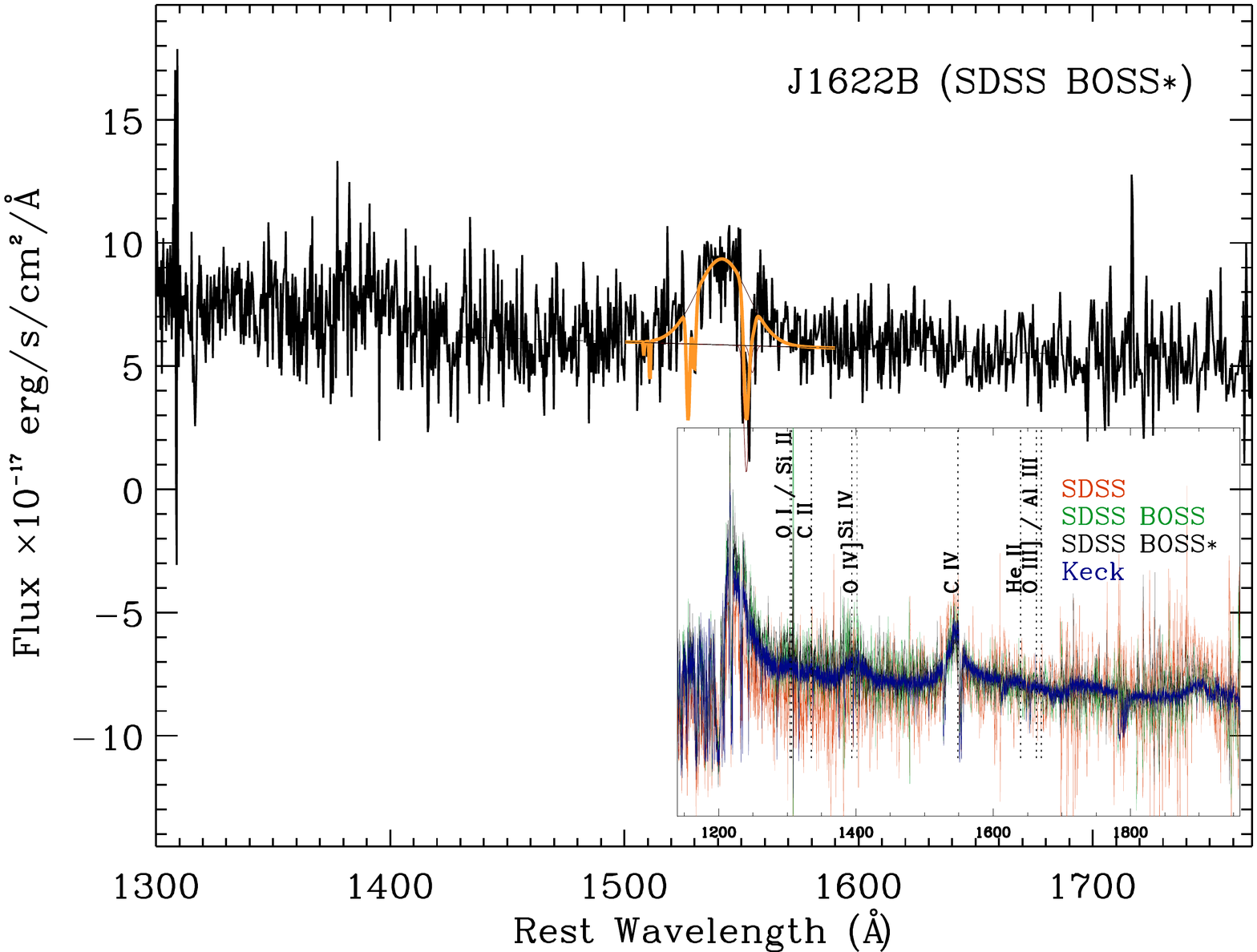}
\vskip0.4cm
\includegraphics[width=8.8cm,trim=60 55 0 0,clip,angle=180]{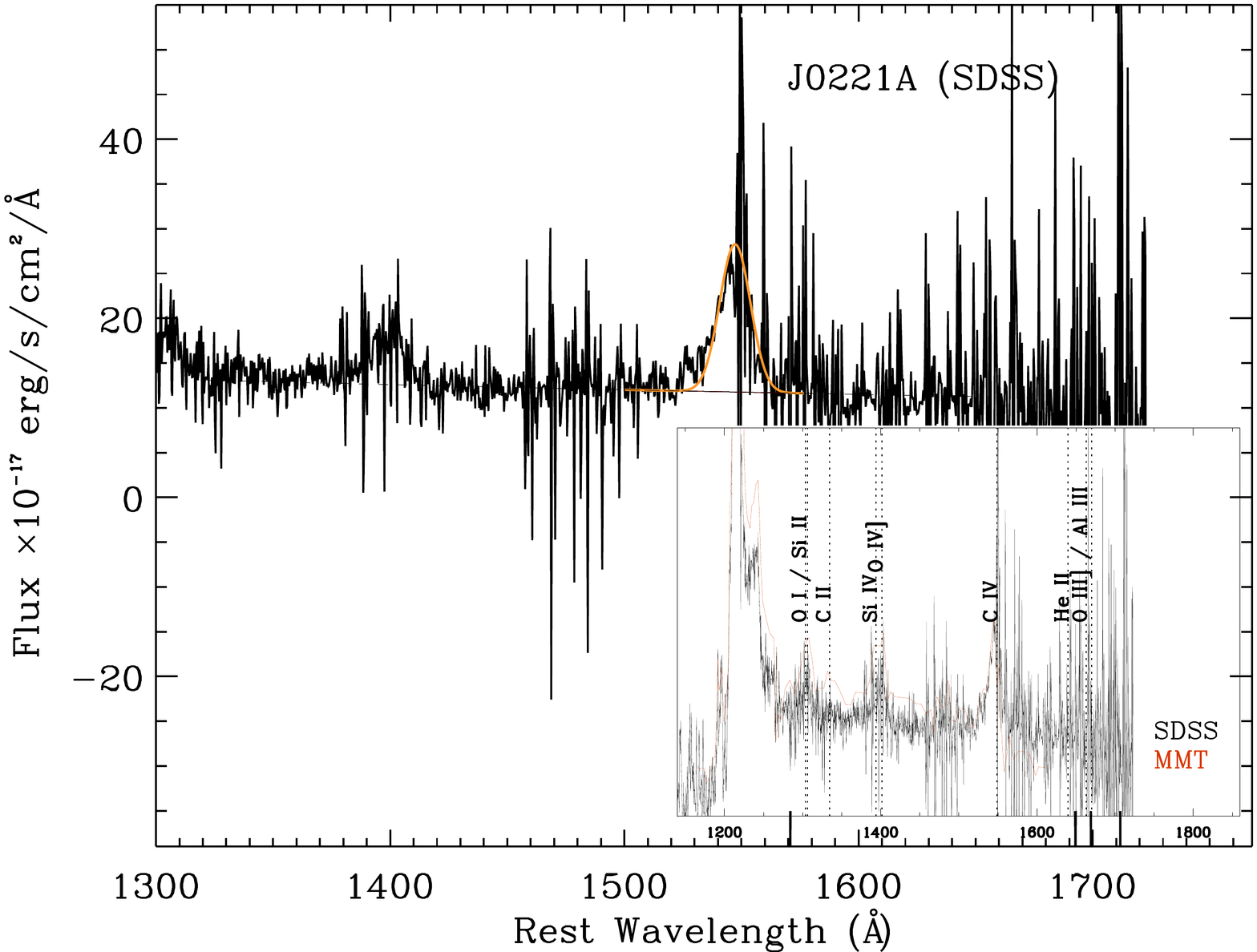}
\includegraphics[width=8.8cm,trim=85 55 0 0,clip,angle=180]{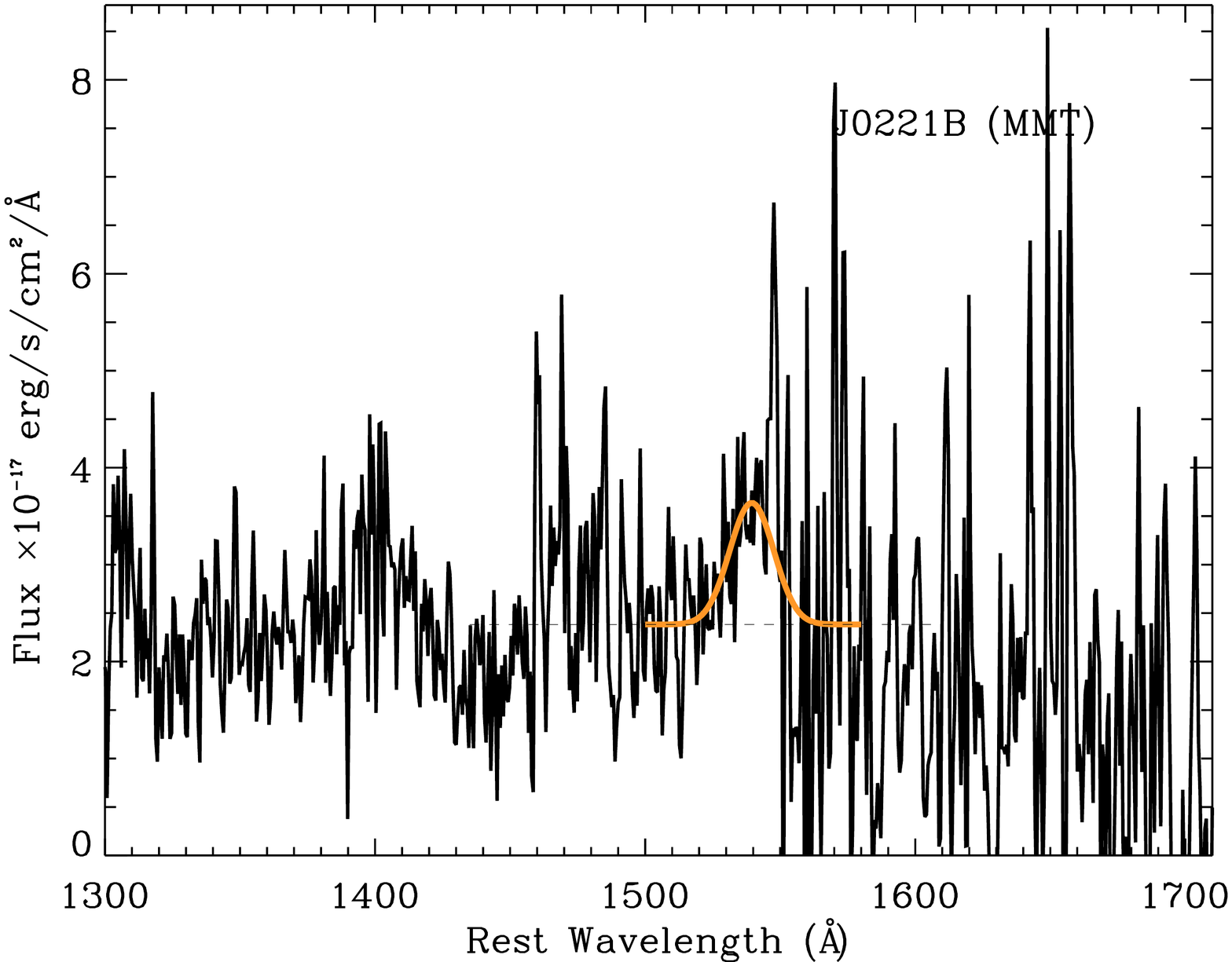}
\caption{Rest-frame UV spectra of the quasar pairs around the \civ\ line showing our fitting method. Superimposed on the spectra are the best-fitting components (see $\S$~\ref{optical}): a power law (dashed line), tracing the continuum emission, and  Gaussian profiles (black solid curves), reproducing the \civ\ emission line. The orange solid curve represents the sum of all components, including a set of negative Gaussian profiles which take into account the BAL contribution when needed. In the insets, we show the various spectra of each quasar (when available) collected from archival data, which allowed us to prove the absence of significant luminosity variability, with the exception of J1307A. For J1307A and J1622B, we selected the SDSS BOSS spectra closer to the \xray\ observation. 
For J0221B, the region at wavelength of 1545\AA\ has been excluded from the fit, because of the strong sky-subtraction residuals affecting the \civ\ line profile (see Fig.~3 of \citealt{mcgreer16}).
Vertical dotted lines mark the wavelength of UV emission lines in the range $1150-1950$\AA.}
\label{optical_spectrum}
\end{figure*}

Four out of the six QSOs have been observed with different facilities (Keck, Magellan, MMT), while two have been observed two/three times as part of different SDSS programs. We used these spectra to check for the presence of spectral variability and the reliability of flux calibration.\footnote{We note that the three SDSS spectra of J1622B, taken in the period 2005--2012, do not show any variation in spectral fluxes and shapes. The Keck spectrum, taken in 2008, is instead several times fainter. We interpret this disagreement as due to wrong flux calibration and chose to re-normalize this spectrum to the SDSS fluxes. Similar differences have been found also between the Keck spectrum of J1622A and its SDSS magnitudes, and for J1307A; also in these cases we re-normalized the Keck spectrum to SDSS fluxes. 
Variability is observed in J1307A, for which four spectra are available, taken in the period 2002--2011 (the latest being the BOSS spectrum, analysed in this paper; see Fig~\ref{optical_spectrum}).}
SDSS or SDSS-III Baryon Oscillation Spectroscopic Survey (BOSS; \citealt{dawson13}) spectra have been preferentially used to derive SMBH properties, because they typically have higher SNR (especially the BOSS ones). 
However, we analysed all available spectra to check the reliability of our results. 

Before modelling the \civ\ line, for all sources but J0221B we subtracted the continuum emission fitting a power law at both sides of the ionized carbon line (in the two windows at 1330\AA\ and 1690\AA). 
The continuum in J0221B was instead modeled with a constant, given the low SNR. Then, we used a multi-Gaussian approach to best reproduce the total profile (see Fig.~\ref{optical_spectrum}). We used up to three Gaussian profiles for the emission component. Negative narrow (broad) Gaussian lines with FWHM of few hundreds (thousands) of km/s have also been taken into account to reproduce narrow (broad) absorption line systems (NAL and BAL, respectively). Emission redwards of $\approx$1600\AA\ can hardly be considered to be associated with \civ\ line \citep{fine10} and was not taken into account in our analysis. Moreover, iron emission contamination (FeII and FeIII; \citealt{vestergaard01}) in the region around \civ\ line is expected to be negligible \citep{shen08, shen11b} and was not fitted. 

Because of the complex shape of \civ\ emission (e.g., \citealt{gaskell82}) and the issues regarding the presence and the treatment of \civ\ narrow line region emission in deriving black hole masses (see detailed discussion in \citealt{shen11b}; see also \citealt{dietrich09} and \citealt{assef11}), we derived non-parametric widths following the prescriptions of \cite{shen11b}, i.e. the FWHM of the best-fitting model profile. When NAL and BAL systems affect the \civ\ profile, we considered the total profile obtained from the only positive Gaussian components.  

We derived black hole masses using the virial theorem and the broad-line region (BLR) radius -- luminosity relation \citep{vestergaard06}. 
The reliability of \civ\ virial mass estimator is, however, strongly debated. On the one hand, \civ\ scaling relation is still based on very few measurements \citep{kaspi07, saturni16, park17}. On the other hand, carbon line is often associated with blueshifted wings likely due to outflows (e.g. \citealt{richards11}); this non-virial component may affect the measurements of the total profile width and bias the black hole mass estimates. Empirical corrections have been proposed to reduce the bias \citep{denney12, park13, coatman17} .

In Table~\ref{optical_results} we report the range of values that can be derived using the relations by \cite{shen11b} (see their Eq.~2), \cite{denney12} (Eq.~1), \cite{park13} (Eq.~3) and \cite{coatman17} (Eq.~6). 
The upper boundary value of each interval is usually obtained using the \cite{shen11b} prescriptions. 
The uncertainties in these measurements are dominated by the intrinsic scatter ($\approx0.3$~dex,  \citealt{park17, denney12, vestergaard06}) in the single-epoch calibrations, which is much larger than the uncertainties ascribed to the quality of the analysed spectra ($\approx0.01-0.05$ dex for the highest SNR spectra). 
To derive BH mass from \cite{coatman17} formula, we measured \civ\ offsets using systemic redshift when the low-ionization narrow line \cii\ is present in the spectra (see Table~\ref{optical_results}). 
For J1622A and J1622B, instead, we maximised the correction taking advantage of the anti-correlation between \civ\ velocity offset and EW (e.g., \citealt{coatman17}): starting from the SDSS DR7 QSO catalog of \cite{shen11b}, for each of our targets we selected an SDSS subsample with similar \civ\ EW (within the errors). Then,  we used these sources to construct a \civ\ velocity offset distribution and, from this, we derived the third quartile value. These velocities have been used to maximise the \cite{coatman17} BH mass correction of J1622A and J1622B.

Bolometric luminosity have been derived from the 1350\AA\ luminosity, extrapolating the continuum power law to short wavelengths. We adopted a bolometric correction of 3.8, valid for luminous SDSS quasars and characterized by very limited scatter (see \citealt{richards06}). For J0221B, \civ\ line is at the edge of the spectral range covered by MMT, where the transmission is very low. In this case,  
the continuum luminosity has been derived through a modelling with a constant. 
The main quasar parameters from the analysis of the optical data are reported in Table~\ref{optical_results}. 

\begin{figure*}
\centering
\includegraphics[width=0.68\columnwidth,angle=0]{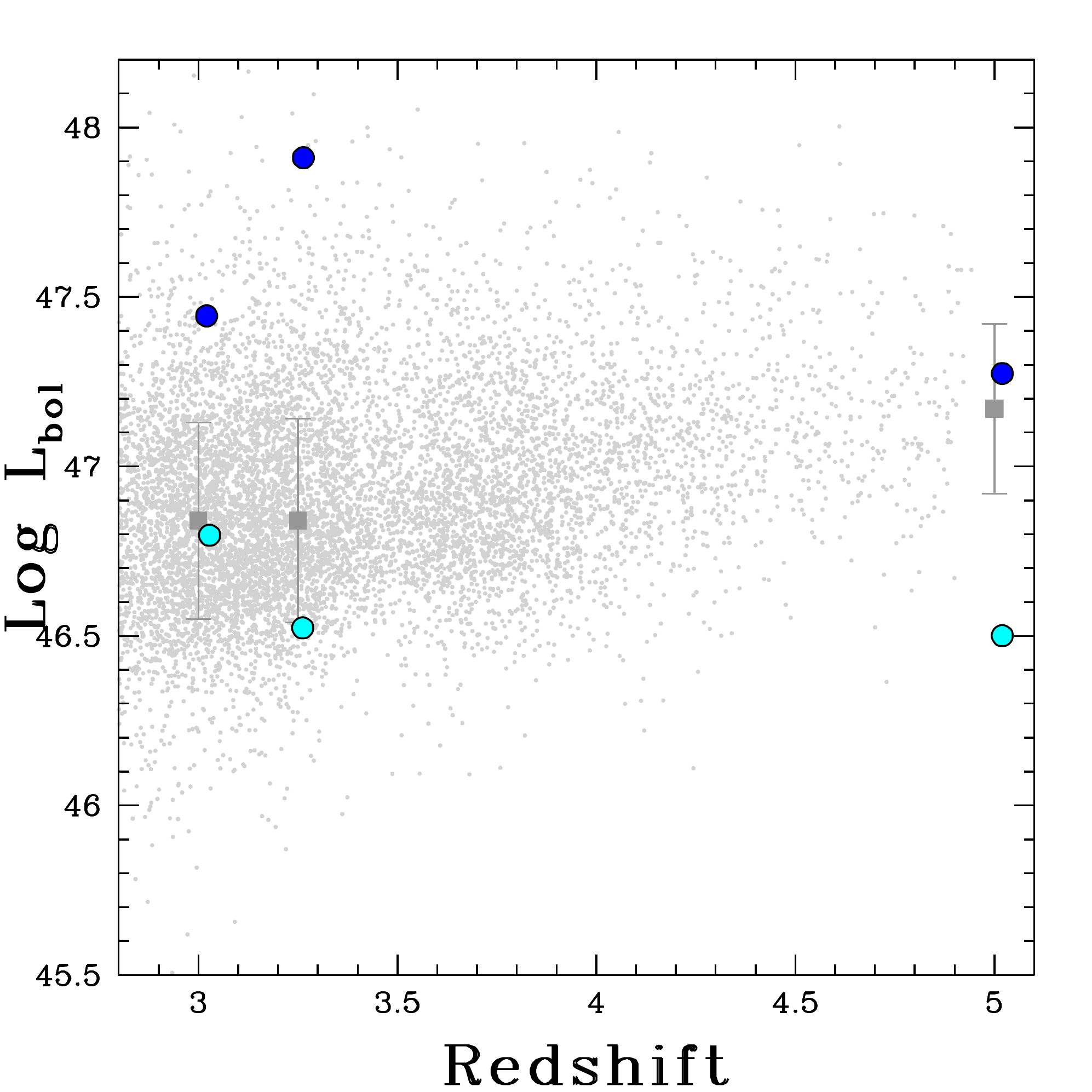}\hfill
\includegraphics[width=0.68\columnwidth,angle=0]{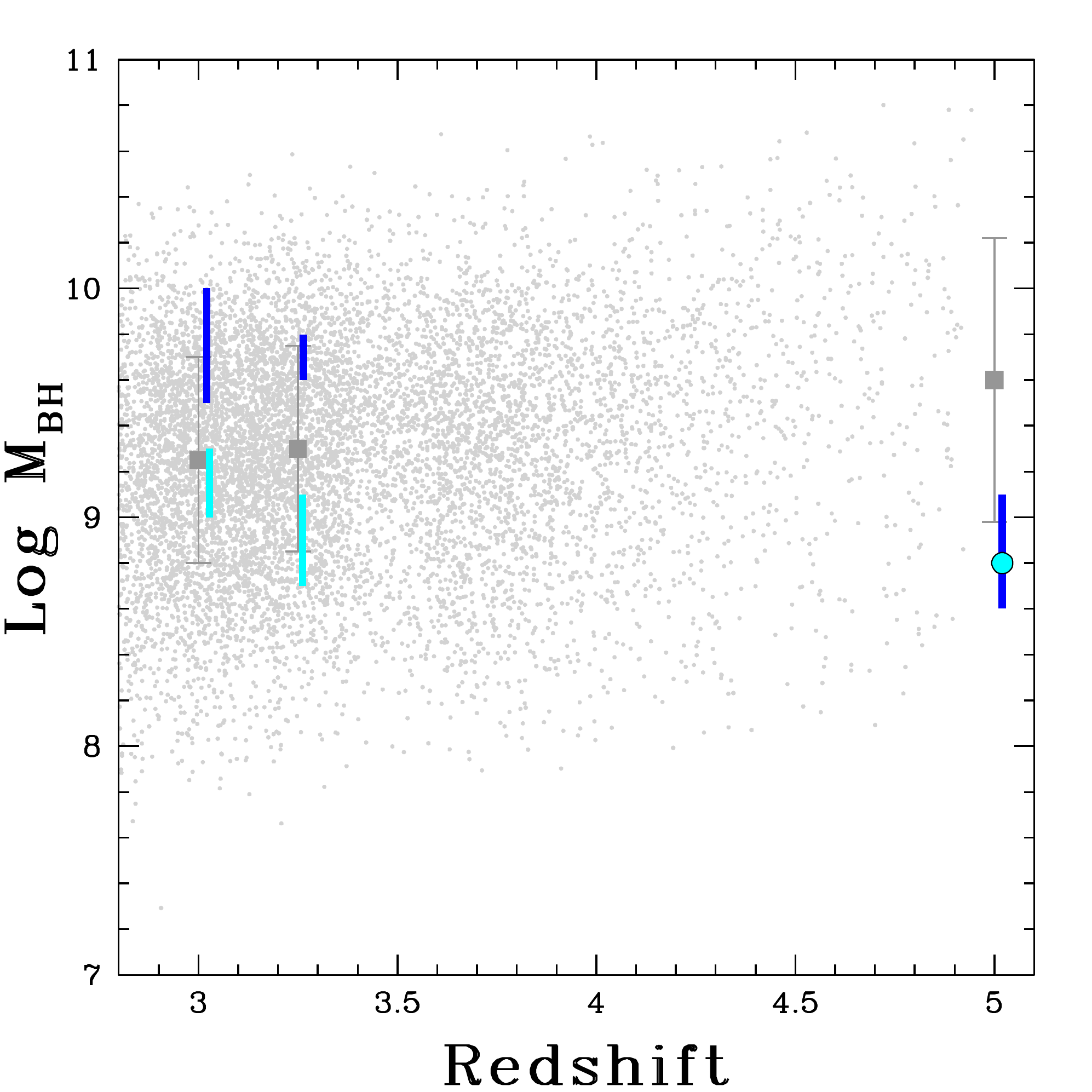}\hfill
\includegraphics[width=0.68\columnwidth,angle=0]{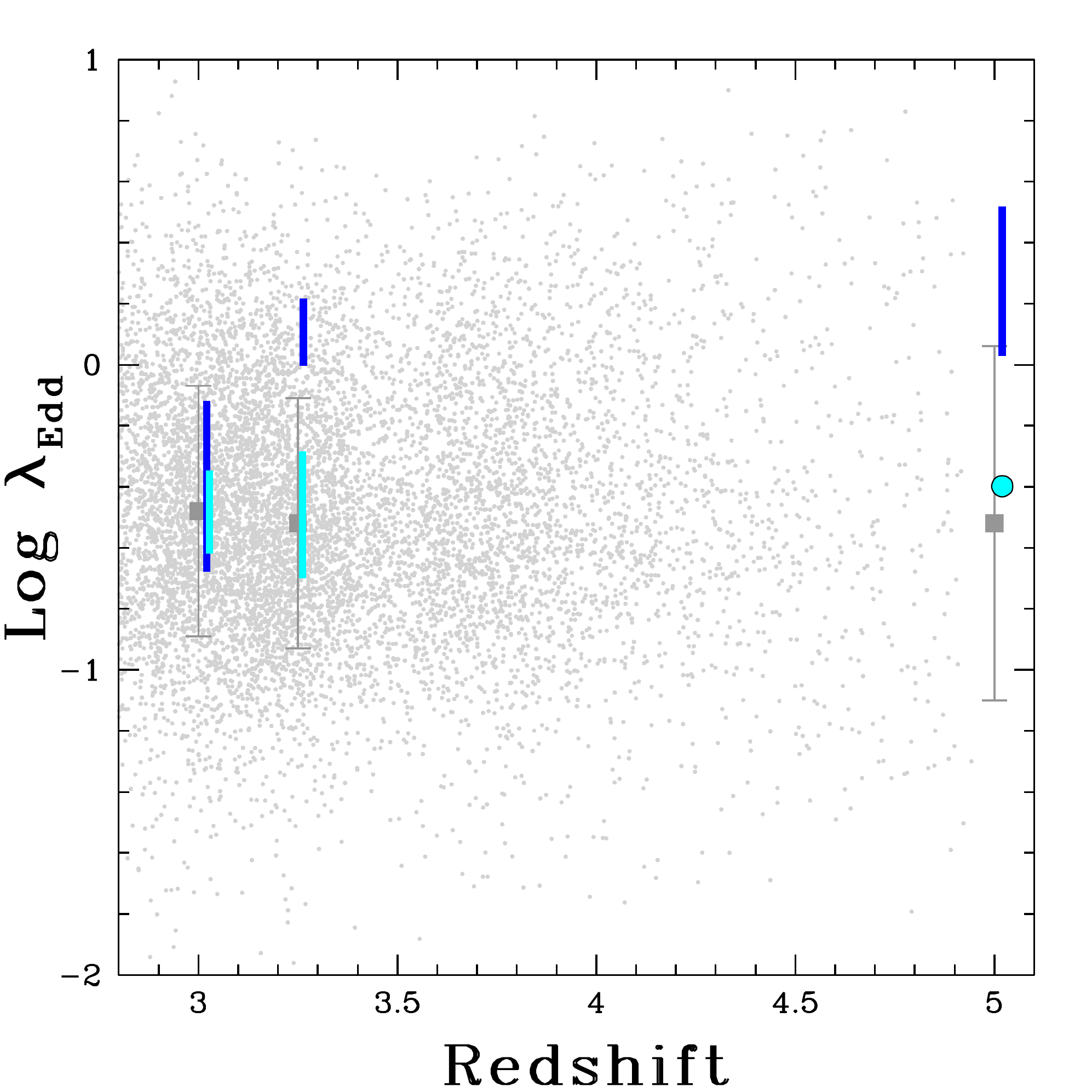}
\caption{Comparison of the properties of the three quasar pairs of the current work vs. SDSS quasars (from the compilation of \citealt{shen11b}, updated as Nov. 2013, small grey dots): bolometric luminosity vs. redshift (left panel), black hole mass vs. redshift (middle panel) and Eddington ratio vs. redshift (right panel). The grey squares represent the median values from the SDSS quasars in redshift ranges encompassing/close to those of our sources, while the error bars show the 1-$\sigma$ dispersion. The A (B) quasar components of our sample are shown as blue (cyan) symbols and bars; the black hole mass and Eddington ranges (excluding J0221B) are computed as described in $\S$\ref{optical}, assuming negligible error on the derived bolometric luminosity (see \citealt{richards11}).
}
\label{comparison_plots}
\end{figure*}

\section{Properties of the quasar pairs at high redshift versus those of SDSS quasars}
\label{comparison}
Although the sample of quasar pairs at high redshift presented in this work is limited, we made an attempt to verify whether, to first order, their properties strongly differ from those of luminous isolated quasars in the same redshift ranges. The results should give an idea on whether AGN in pairs have peculiar properties, maybe ascribed to the presence of the active companion on scales of several tens of kpc.  

{\it In primis}, we have computed the source optical-to-X-ray power-law slope, \aox, which is a measure of the relative importance of the emission disc vs. hot corona. It is defined as 
$\alpha_{\rm ox}=\frac{\log(f_{\rm 2~keV}/f_{2500~\mbox{\rm \scriptsize\AA}})}{\log(\nu_{\rm 2~keV}/\nu_{2500~\mbox{\rm \scriptsize\AA}})}$, 
where $f_{\rm 2~keV}$ and $f_{2500~\mbox{\scriptsize \rm \AA}}$ are the 
rest-frame flux densities at 2~keV and 2500~\AA, respectively. These quantities have been obtained from the 
available \xray\ spectra and optical photometry using the method outlined in \cite{vignali03}. 
The 1-$\sigma$ errors on \aox\ are typically of $\approx0.05-0.15$ once the uncertainties on the \xray\ and optical values are considered (with the former contributing most). 
The \aox\ values of our non-BALQSOs ($\approx-1.4/-1.8$) are within the range expected for quasars once the \aox\ vs. 2500~\AA\ luminosity anti-correlation (and relative scatter) is taken into account (e.g, \citealt{vignali03, shemmer06, just07, lusso10, nanni17, martocchia17}). The sources deviating most from the expected values, assuming the \cite{just07} relation, are those which are classified as BALQSOs in the present sample ($\approx-2.0/-2.2$). This is not unexpected, since steeper \aox\ (i.e., fainter soft X-ray emission at a given optical luminosity) are typically observed in absorbed sources as BALQSOs (e.g., \citealt{gibson09}). 

For a more exhaustive comparison of our sample vs. high-redshift quasars, we have drawn three subsamples from the compilation of SDSS quasars published by \cite{shen11b}, after checking that none of them has an active companion up to $\approx$100~kpc; in particular, we used the updated values of bolometric luminosities, black hole masses and Eddington ratios\footnote{$\lambda_{\rm Edd}=L_{\rm bol}/L_{\rm Edd}$, where L$_{\rm bol}$ is the bolometric luminosity and L$_{\rm Edd}$ is the Eddington luminosity ($1.3\times10^{38}~(M_\mathrm{BH}/M_\odot$)~\lum).} from the most recent (as Nov. 2013) version of the Shen catalog. For the J1307 quasar pair, we selected 1047 quasars at $z=2.95-3.05$, while for the J1622 system, 965 quasars were extracted at $z=3.20-3.30$. More challenging is to find an SDSS parent sample at $z\approx5$: to this goal, we used the values from \cite{shen11b} for 55 quasars at $z=4.8-4.95$ (no quasar is present at higher redshift). 
The bolometric luminosities, black hole masses and Eddington ratios of the three subsamples are shown as little grey points in Fig.~\ref{comparison_plots} (left, middle, and right panels, respectively); the median values and the dispersions of the distributions are reported as large grey squares and corresponding error bars. The values derived in $\S$\ref{optical} for the three quasar pairs of this work are shown as blue and cyan symbols and bars for the A and B components of each pair, respectively (where bars take into account the range of black hole masses, hence of Eddington ratios, as described in $\S$\ref{optical}). 

Starting from the derived bolometric luminosities (left panel in Fig.~\ref{comparison_plots}), we note that the A components of the $z\approx3-3.3$ quasar pairs are well outside (and above) the median values of SDSS quasars; 
about the B components, J1622B is on the lower end of the 1$\sigma$ distribution, while J0221B is well below. 
However, the sampling of $z\approx5$ quasars in \cite{shen11b} compilation is, as already stated, particularly scarce. 
Our results still hold if we compare the two BALQSOs, J1307B and J1622A, with the BALQSOs in \cite{shen11b} (166 and 135 at $z\approx3$ and $z\approx3.3$, respectively). 

For what concerns the black hole masses (plotted as bars to take into account the range reported in Table~\ref{optical_results}, except for J0221B, for which only the Shen prescription has been adopted), the A components of our $z\approx3-3.3$ pairs lie in the upper envelope of the SDSS BH mass distribution (middle panel of Fig.~\ref{comparison_plots}), although within the dispersion. 
J1622A is still located ``high" in the distribution if the comparison is done against the BALQSOs at $z\approx3.3$. 
Among the B components, the only ``outlier" (but see the note of caution above) is J0221B, having a black hole mass apparently lower than the median of the distribution at $z\approx5$. If we assumed an uncertainty for its black hole mass similar to that of its A companion, J0221B would be most likely within the dispersion of the distribution. 

All these results translate into the right panel of Fig.~\ref{comparison_plots}, where the Eddington ratio (which is essentially the ratio of the quantities plotted in the y-axes in the other two panels, with the bars showing the range of black hole masses, hence Eddington luminosities) is reported for the three quasar pairs vs. SDSS quasars. Both J1307 A and B quasars are within the distribution; J1622A is above the median value (because of its very extreme bolometric luminosity) and a similar result is found for J0221A (in virtue of its relatively low BH mass). 

The $z\approx3.2$ quasar luminosity function (QLF) presented by \cite{masters12} allows us to place our 
$z\approx3-3.3$ quasar pairs in a broader context: all of them populate the bright end of 
the QLF (on the basis of their rest-frame 1450\AA\ absolute magnitudes, M$_{1450}$), with J1622A being on one extreme (M$_{1450}=-28.6$), and the companion, J1622B, being close to the knee of the QLF (with M$_{1450}=-25.4$). 

As already said, the comparison of the properties of J0221 quasars with those of quasars at similar redshift is more challenging. To understand how (a)typical J0221 quasars' properties are (besides the uniqueness of this quasar pair at such high redshift; see the extended discussion in $\S4.2$ of \citealt{mcgreer16}), we start computing their M$_{1450}$ from the photometry reported in \cite{mcgreer16}, obtaining M$_{1450}\approx-26.9$ and M$_{1450}\approx-25.0$ for J0221A and J0221B, respectively. Using, as reference, the work by \cite{mcgreer13}, where the $z\approx5$ QLF  is derived by measuring the bright end with the SDSS and the faint end with the two-magnitude deeper SDSS data in the Stripe~82 region, J0221A clearly lies among the optically most luminous quasars and J0221B is around the mean. If we focus at $z=4.9-5.1$, J0221A is actually brighter by $\approx0.3$ magnitude than the most luminous SDSS quasar reported in \cite{mcgreer13} in this redshift range, while J0221B is around the median value. 

As previously stated, the main limitation of the current work lies on the small number of quasar pairs at high redshift with well defined multi-wavelength properties, including sensitive \xray\ coverage to establish the level of nuclear activity. Extending the present sample would be important for a more solid assessment of the issues related on how the accretion-related activity in pairs is above the median value found for quasars not in pairs at similar redshifts. Such an extension, which could start from e.g. the compilation of H10, would be quite expensive in terms of observing time even with the present-day sensitive \xray\ facilities (\chandra, \xmm), as shown in the current work, where a good/moderately good description of quasar \xray\ properties (continuum slope, column density, intrinsic luminosity) has required $\approx$~65--80~ks exposures. We also note that in Cycle~15 the effective area of \chandra, especially in the soft band, was much higher than the current one (by about 60~per~cent in the 0.5--7~keV band). 
Having said that, it seems that the presence of an active companion does not provide a substantial increase in terms of accretion (namely, Eddington ratio), and this appears to be consistent with the lack of signs of interactions in the available optical images of our target fields. 
The possible exception is represented by J1622A at $z\approx3.3$ (and, more tentatively because of the limited comparison sample, of J0221A at $z\approx5$). 
Focusing on J1622A, we computed the probability of extracting a bolometric luminosity higher than that of J1622A (8.14$\times10^{47}$~\lum; see Table~\ref{optical_results}) by chance from the sample of 965 SDSS quasars at $z=3.20-3.30$ reported above. We note that in this ``parent" sample only five objects have a bolometric luminosity above that of J1622A and, in the same redshift range, there are three quasar pairs (hence six objects) in the H10 original sample. Thus we randomly extracted (one million trials) six quasars from the 965 SDSS quasar sample and obtained a bolometric luminosity above 8.14$\times10^{47}$~\lum\ in 3.1~per~cent of the cases. Therefore, our result of an increase of activity in J1622A due to the presence of an active companion has a not-negligible possibility to be obtained by chance (i.e., our result is significant at the $\approx2.3\sigma$ level).

\section{Summary of the results and conclusions}
\label{summary}
In this paper we have presented, for the first time, the \xray\ and optical properties of three quasar pairs at $z\approx3$ and $z\approx3.3$ (selected from the SDSS and H10 work; their projected separations are 65 and 43~kpc, respectively) and $z\approx5$ (selected from the CFHTLS, with a separation of 136~kpc; \citealt{mcgreer16}). The goal of this work, which benefits from the sensitivity and spatial resolution of \chandra\ (two 65-ks proprietary observations for the $z\approx3-3.3$ systems) and \xmm\ (one ``cleaned"  $\approx$80-ks archival pointing for the $z\approx5$ system), is to provide some, though preliminary, indications on how accretion onto SMBHs and subsequent emission of radiation is possibly enhanced by the presence of an active companion on tens of kpc scales with respect to isolated quasars at similar redshifts. This is a first attempt, conducted on an admittedly small sample, to tackle the issues related to dual quasar activity at high redshift, whereas most of the studies carried out in recent years, at much lower redshifts ($z\simlt0.2$) and luminosities, indicate enhanced BH and star-formation activity at close (several kpc) separations (e.g., \citealt{ellison11}). 
In the following we summarize the main results: \\
$\bullet$ 
Both quasar components of the three pairs are detected with relatively good photon statistics to allow us to derive the column density and intrinsic \xray\ luminosity ($\approx$~a~few~10$^{44}-10^{45}$~\lum\  for all quasar pairs). \\
$\bullet$
Absorption of the order of a few $\times10^{23}$~cm$^{-2}$ has been clearly detected in the two quasars of the systems at $z\approx3.0-3.3$ (J1307B and J1622A) whose optical spectra are characterized by the presence of broad absorption features (likely associated with outflowing winds). This is not unexpected based on works on isolated BALQSOs at high redshift. Furthermore, tentative absorption of a few $\times10^{22}$~cm$^{-2}$ has been found in J1307A, possibly suggesting that gas was destabilized in the central region of this object by the encounter with the companion. \\
$\bullet$
Using the information obtained from the analysis of the optical spectra of all quasar pairs, we derived bolometric luminosities, black hole masses and Eddington ratios. In particular, comparing their Eddington ratios vs. those of luminous SDSS quasars in the same redshift intervals we find that only one source, J1622A, lies significantly above the distribution of SDSS quasars at $z\approx3.3$. One possible explanation of this result is that the high level of activity of J1622A (a bolometric luminosity of $\approx8.1\times10^{47}$~\lum\ for a $\log\ (M_{\rm BH}$/\msun)=9.6--9.8, about an order of magnitude higher than the bolometric luminosity of SDSS quasars at the same redshift) may be linked to the presence of the companion quasar. 
A statistical analysis of this result indicates a 3.1~per~cent possibility that it is obtained by chance. 

The results obtained for our sample of high-redshift dual quasars rely upon the limited number of quasars analysed thus far and the limited quality of \xray\ spectra for some of them. To draw more solid conclusions in terms of properties of dual quasars at high redshifts and how these differ from those of  quasars not in pairs at similar redshift, one would need deeper investigation with \chandra\ and \xmm\ starting from the entire sample of H10, in which further six quasar pairs at separations $<100$~kpc are available (25 up to 650~kpc); another possibility, more suitable for \chandra\ given the low separations ($<6.3$\arcsec), consists in observing the sample presented by \cite{eftekharzadeh17}. This project plan would match the large-project requirements for \chandra\ and \xmm\ calls and, coupled with deep optical/near-IR observations, would allow also a study of source overdensities in these quasar fields. 

A similar investigation, aimed at finding companions in our quasar fields, can be carried out 
in the submillimer by \alma\ and {\it NOEMA}, using the molecular high-J CO transitions and the [\cii] lines as tracers of molecular and ionized gas and, possibly, the spectral scanning mode. With respect to other wavelengths (including X-rays, at least in part), submillimeter/far-infrared observations have the advantage of allowing potential detection, at high redshift, of very obscured AGN, besides galaxy companions (e.g., \citealt{carniani13, decarli17}). 
While galaxies at $z\approx3.3$ can be found using the CO(4-3) transition (Band~3), systems at $z\approx3.0$ and $z\approx5.0$ in our quasar fields can be detected using the CO(5-4) line (Band~4 and 3, respectively). A similar search for companions would also benefit from the [\cii] line -- particularly strong in star-forming systems -- with observations in Band~8 and Band~7 for $z\approx3-3.3$ and $z=5.0$ systems, respectively.

\section*{Acknowledgements}
The authors thank the referee for her/his useful comments and suggestions. They also thank E. Lusso and F. Vito for useful discussions, F. Pozzi for help with IDL, and I. McGreer for providing us with the optical spectra of J0221 quasars. A special thank to all the members (and friends) of the Multiple AGN Activity (MAGNA) collaboration. 
Support for this work was provided by the National Aeronautics and Space Administration through Chandra Award Number GO4-15105 issued by the Chandra X-ray Observatory Center, which is operated by the Smithsonian Astrophysical Observatory for and on behalf of the National Aeronautics Space Administration under contract NAS8-03060.







\bsp	
\label{lastpage}


\end{document}